\documentclass[prb,aps,twocolumn,amsdtx,showpacs]{revtex4}
\usepackage{graphicx}
\usepackage{amsmath}
\usepackage{amssymb}
\usepackage{color}
\pdfoutput=1

\newcommand{\bleq}{\ifpreprintsty
                   \else
                   \end{multicols}\vspace*{-3.5ex}{\tiny
                  \noindent\begin{tabular}[t]{c|}
                  \parbox{0.493\hsize}{~} \\ \hline \end{tabular}}
                   \fi}
\newcommand{\eleq}{\ifpreprintsty
                 \else
                   {\tiny\hspace*{\fill}\begin{tabular}[t]{|c}\hline
                    \parbox{0.49\hsize}{~} \\
                    \end{tabular}}\vspace*{-2.5ex}\begin{multicols}{2}
                    \fi}
\newcommand{\bcols}{\ifpreprintsty\else\begin{multicols}{2}\fi}
\newcommand{\ecols}{\ifpreprintsty\else\end{multicols}\fi}

\newcommand \beq  {\begin{equation}}
\newcommand \eeq  {\end{equation}}
\newcommand \bea {\begin{eqnarray} }
\newcommand \eea {\end{eqnarray}}

\begin{document}

\title{Antiferromagnetic Ising model on the sorrel net: a new frustrated corner-shared triangle lattice}

\author{John M. Hopkinson}
\affiliation{Brandon University, Brandon, Manitoba, Canada, R7A 6A9}
\author{Jarrett J. Beck}
\affiliation{Brandon University, Brandon, Manitoba, Canada, R7A 6A9}
\date{\today}

\begin{abstract}
 We study the antiferromagnetic classical Ising (AFI) model on the sorrel net, a $\frac{1}{9}$th site depleted and $\frac{1}{7}$th bond depleted triangular lattice.  Our classical Monte Carlo simulations, verified by exact results for small system sizes, show that the AFI model on this corner-shared triangle net (with coupling constant $J_1$) is highly frustrated, with a residual entropy of $\frac{S}{N}$= 0.48185$\pm$0.00008.  Anticipating that it may be difficult to achieve perfect bond depletion, we investigate the physics originating from turning back on the deleted bonds ($J_2$) to create a lattice of edge-sharing triangles.  Below a critical temperature which grows linearly with $J_2$ for small $J_2$, we identify the nature of the unusual magnetic ordering and present analytic expressions for the low temperature residual entropy. We compute the static structure factor and find evidence for long range partial order for antiferromagnetic $J_2$, and short range magnetic order otherwise.  The magnetic susceptibility crosses over from following a Curie-Weiss law at high temperatures to a low temperature Curie law whose slope clearly distinguishes ferromagnetic $J_2$ from the $J_2 = 0$ case.  We briefly comment on a recent report \cite{keene} of the creation of a $\frac{1}{9}$th site depleted triangular lattice cobalt hydroxide oxalate.
\end{abstract} 

\maketitle

\section{Introduction}
The study of geometrically frustrated  magnetic materials provides a fruitful path to the discovery of exotic physics.  In such materials, competing interactions between magnetic moments due to their localization on a highly symmetric lattice structure lead to large numbers of quasi-degenerate states.  Interesting physics resulting from such near degeneracies on frustrated lattices is thought to include: heavy fermion behavior in the spinel LiV$_2$O$_4$\cite{urano}, evidence for the existence of magnetic monopoles in the spin ice pyrochlores Dy$_2$Ti$_2$O$_7$\cite{Bramwell} and Ho$_2$Ti$_2$O$_7$\cite{Fennell1}, non-Fermi liquid behaviour in the distorted kagome based FeCrAs\cite{Wu}, spin liquid behavior in hyperkagome Na$_4$Ir$_3$O$_8$\cite{okamoto} and triangular lattice based $\kappa$-(BEDT-TTF)$_2$Cu$_2$(CN)$_3$\cite{shimuzu}, and unusual yet to be understood Kondo physics in the pyrochlore Pr$_2$Ir$_2$O$_7$\cite{nakatsuji}.  

While in traditional magnetic materials a two-dimensional net might arise in a layered lattice structure, a growing body of work has focused on designing model systems to test models of strongly correlated electron systems.  The study of two-dimensional frustrated systems has been augmented by the creation of artificial spin ice systems\cite{Wang,Qi}, where people have shown that it is possible to lithographically etch ferromagnetic islands in patterns that can be easily modeled.  From a theoretical perspective, the dipolar interactions crucial to such spin ice physics converge quickly in two dimensions, so that it is feasible to sum the dipolar interactions in real space.  Recent work on triangular optical lattices\cite{struck}, magnetic colloids\cite{steinbach} and macroscopic spin ice\cite{melladomacro} have generated further interest in two dimensional frustrated spin models. 

\begin{figure}[here]
\label{figure1}
\includegraphics[scale=0.55]{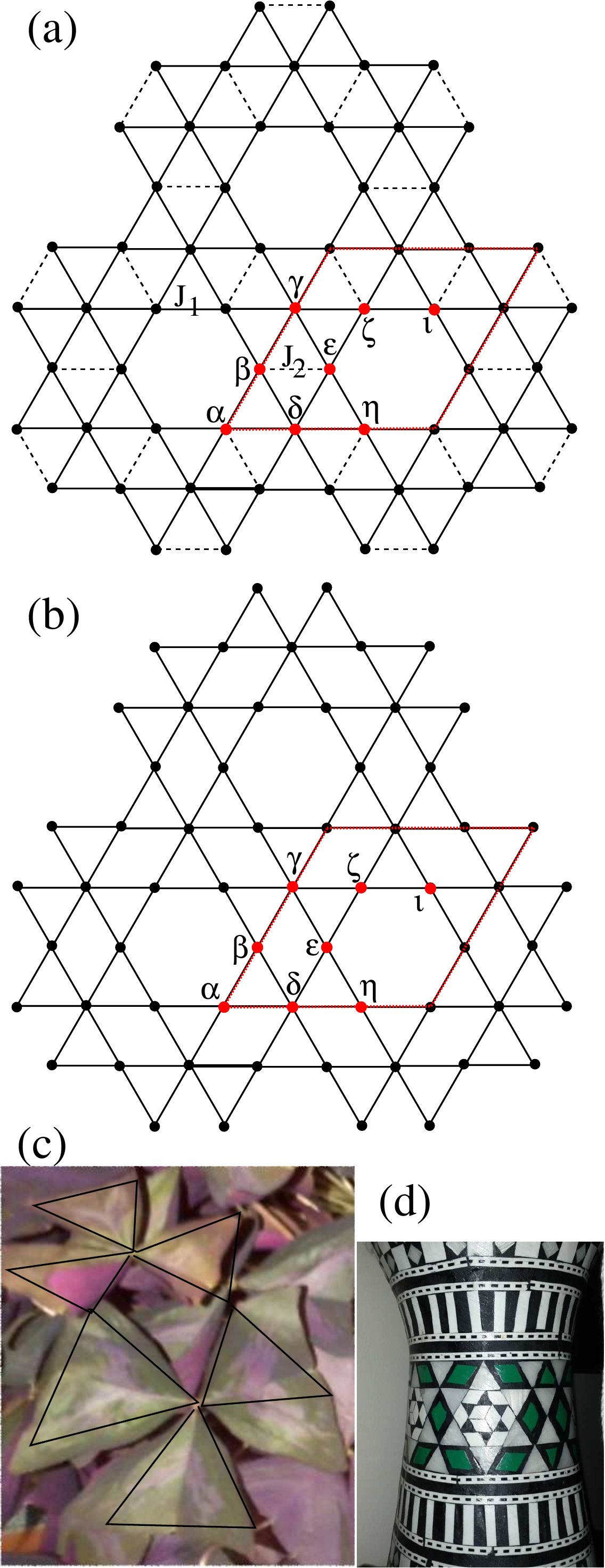}
\caption{(Color online)(a) Selective removal of 1/9th of the sites of the triangular net gives an new edge-shared net.  The further removal of 1/7th of the bonds (indicated by dashed lines labeled $J_2$) results in the sorrel net shown in (b).  There are 8 sites in the unit cell, six of which are 4-coordinated $\{\alpha,\beta,\epsilon,\zeta,\eta,\iota\}$, and 2 of which are six coordinated $\{\gamma,\delta\}$.  (c) A photograph of oxalis regnelli triangularis, commonly known as wood sorrel with lines drawn atop it to guide the eye. (d) a common pattern on a doumbek.
\label{figure1} }
\end{figure}

The Ising model on the two dimensional triangular net has been studied since the 1950's\cite{Wannier} and been shown to possess a large spin degeneracy as evidenced by a residual ground state entropy per spin of $\frac{S}{N} = 0.323$\cite{Nagai}. The simplest frustrated depletion of this edge-shared bipartite net removes every fourth spin in a regular fashion, substituting a non-magnetic atom for every second atom along each of the basis vectors, to create the kagome net.  The kagome net possesses a larger ground state degeneracy for the Ising model than the triangular net due to the corner-sharing nature of its bonds, as shown by an enlarged residual entropy per spin of $\frac{S}{N}= 0.50183$\cite{Kano}.  After many years of experimental searches, the kagome net has been experimentally realized both in quantum spin systems\cite{hiroi,shores,colman,schluter,okamotohiroi,ono} and artificial spin ice systems\cite{Qi}.  The triangular kagome net is a $\frac{7}{16}$th depleted triangular net, removing every fourth spin along the basis vectors and their neighboring hexagons.  Once again, the antiferromagnetic Ising model features a large residual entropy per spin, $\frac{S}{N} = \frac{1}{9}\ln(72)\approx 0.475185$.\cite{loh}  The maple leaf net\cite{betts} is a $\frac{1}{7}$th depleted triangular net, removing a triangular lattice formed by advancing along one basis vector by two units and a second by one unit in a manner similar to a knight in chess.  The Ising model on this edge-sharing net, which has coordination number 5 in contrast to the triangular (6) and triangular kagome and kagome (4) coordinations, appears to have not yet been studied.  Experimental candidates for both the triangular kagome\cite{mekata} and maple leaf\cite{cave,hawthorne,fennell} nets have been found.

In this paper we treat the Ising model on the sorrel net, a  1/9th site depleted and further 1/7th bond depleted triangular corner-shared triangle net as described below in Sect. \ref{section1}.  By means of Monte Carlo simulations, we find that the classical antiferromagnetic Ising model on this net has a finite residual entropy per spin in the thermodynamic limit of $\frac{S}{N}$ = 0.48185$\pm$0.00008, slightly greater than the Pauling entropy estimate for this net, indicating that the Ising model on this net is more highly frustrated than on the triangular and triangular kagome nets, even approaching that on the kagome net. We calculate the heat capacity, acceptance rate, magnetic susceptibility and static structure factor as a function of temperature to aid the experimental identification of candidate systems.  We investigate the stability of this model to the presence of magnetic interactions on the removed bonds of ferromagnetic or antiferromagnetic sign, and show that for small bond strengths the residual entropy in the spin system approaches that of the sorrel net to quite low temperatures prior to undergoing a secondary transition to a state still possessing a finite residual entropy in the $T\rightarrow 0$ limit.

\section{A new two-dimensional lattice}\label{section1}


In this paper we introduce a 1/9th depleted triangular net as shown in Fig.\ref{figure1}(a) of symmetry P6mm.  One sees that the substitution of non-magnetic atoms (which themselves form a triangular lattice\cite{henley}) for every third atom along each of the basis vectors of the triangular lattice leads to an edge-shared net with sites 6-fold and 5-fold coordinated. Amazingly, the symmetric selective removal of 1/7 of the bonds on this lattice leads to a new highly frustrated {\it{corner-shared}} equilateral triangle two-dimensional net as shown in Fig. \ref{figure1}(b).  This creates a net reminiscent of {\it{oxalis regnelli triangularis}} (commonly known as wood sorrel) leaves as pictured in Fig.\ref{figure1}(c), and for this reason we will henceforth refer to this kagome-inspired net as the sorrel net.  Fig. \ref{figure1} (d) shows a picture of a doumbek.  The corner-shared triangle structure pictured in Fig. \ref{figure1} (b) is often found on the side of these drums, which are traditionally covered by mother-of-pearl, using simple Islamic tiling patterns.

  On the sorrel net the magnetic sites are four and six coordinated, such that one has two types of sites: one connecting two corner-shared triangles, another connecting three corner-shared triangles.  To our knowledge this net has not yet been found in nature.  It is our hope that our identification of this structure as an interesting frustrated lattice will encourage experimental groups to investigate materials naturally forming in this pattern, as has been the case with recent experimental searches for the maple leaf and kagome lattices.  Indeed, one recent study\cite{keene} has managed to achieve a $\frac{1}{9}$th depleted triangular lattice in a magnetic system, although, as we discuss below in Sec. \ref{section5}, the antiferromagnetic order seen may have resulted from additional hexagonal layers in this complex material.

\begin{table}
\begin{tabular}{|l|l|l|l|}
\hline
 Site & Position & Neighboring Sites: $J_1$ & $J_2$ \\
\hline
$\alpha$ & (0,0,0) &$\beta , \delta , \gamma_{ - {\bf{A}}_2} , \eta_{ - {\bf{A}}_1}$ & $\zeta_{-{\bf{A}}_2}$\\ \hline
$\beta$  &  $( \frac{a}{6}, \frac{ \sqrt{3} a }{6}, 0 )$ & $\alpha , \gamma , \delta , \iota_{ - {\bf{A}}_1}$ & $\epsilon$ \\ \hline
$\gamma$ & $( \frac{a}{3} , \frac{ \sqrt{3} a }{3}, 0 )$ & $\beta , \epsilon , \zeta , \alpha_{ + {\bf{A}}_2} , \iota_{ - {\bf{A}}_1} , \eta_{ - {\bf{A}}_1 + {\bf{A}}_2}$ &  \\ \hline 
$\delta$ & $( \frac{a}{3} , 0 , 0 )$ & $\alpha ,\beta , \epsilon , \eta , \zeta_{ - {\bf{A}}_2} , \iota_{ - {\bf{A}}_2}$ & \\ \hline
$\epsilon$ & $( \frac{a}{2} ,\frac{ \sqrt{3} a }{6} , 0 )$ & $\gamma , \delta , \zeta , \eta$ & $\beta$\\ \hline
$\zeta$ & $( \frac{2 a }{ 3 } , \frac{ \sqrt{3} a }{3} , 0 )$ & $\gamma , \epsilon , \iota , \delta_{ + {\bf{A}}_2}$ & $\alpha_{+{\bf{A}}_2}$\\ \hline
$\eta$ &  $( \frac{2 a }{ 3 } , 0 , 0 )$ & $\delta , \epsilon , \alpha_{ + {\bf{A}}_1} , \gamma_{ + {\bf{A}}_1 - {\bf{A}}_2}$ & $\iota_{-{\bf{A}}_2}$\\ \hline
$\iota$ & $( a , \frac{ \sqrt{3} a }{3} , 0 )$ & $\zeta , \beta_{ + {\bf{A}}_1} , \gamma_{ + {\bf{A}}_1} , \delta_{ + {\bf{A}}_2}$ & $\eta_{+{\bf{A}}_2}$\\ 
\hline
\end{tabular}
\caption{  Sites, positions and nearest neighbors for atoms in the unit cell of the sorrel net. When not in the first unit cell neighbors are connected by basis vectors ${\bf{A}}_1 = (a,0,0)$ and ${\bf{A}}_2 = ( \frac{a}{2} , \frac{ \sqrt{3} a }{2} , 0 )$. $J_1 > 0$ is an antiferromagnetic bond between corner shared neighbors while $J_2$ connects the remaining nearest neighbors and is treated in three limits:$\{ -J_1 \leq J_2 < 0, J_2 = 0, 0 < J_2 \leq J_1 \}$. The limit $J_2 = J_1$ corresponds to the antiferromagnetic Ising model on the $\frac{1}{9}^{\text{th}}$ depleted triangular net. }
\end{table}

\section{Model}

  We consider an antiferromagnetic Ising model on the sorrel lattice with the Hamiltonian,
 \begin{equation}
 H = J_1\sum_{\langle ij \rangle}\!\sigma^z_{i}\sigma^z_{j}+J_2\sum_{\langle ik\rangle}\!\sigma^z_{i}\sigma^z_{k} \text{,}\label{equation8}
 \end{equation}
 where $J_1>0$ is an antiferromagnetic interaction between nearest neighbors on the corner-shared sorrel lattice, and $J_2$ is an interaction between nearest neighbor sites that when coupled create an edge-shared net. In our simulations we allow $J_2$ to vary in strength from $-J_1$ to $J_1$.
 
 \section{Method}
 
 We use the metropolis Monte Carlo algorithm to simulate $8\times L\times L$ nets subject to periodic boundary conditions.  Lattice sizes have been varied from $L$ = 1 to 18 on the sorrel lattice with $J_2$ =0 to ensure that the thermodynamic limit has been reached.  At each temperature, prior to averaging, $2\times 10^5$ to $10^6$ Monte Carlo steps have been used to equlibrate the spin system, with each average then taken over a further 2001 to 20001 Monte Carlo steps, where each step corresponds to on average attempting an update on each of the spin sites once.  To simulate the annealing\cite{Kirkpatrick} of an experimental material from high temperatures we have started with the temperature equal to 6000 $J_1$, reducing the temperature by multiplying by a factor of 0.99 for successive measurements.  

\begin{figure}
\includegraphics[scale=0.35]{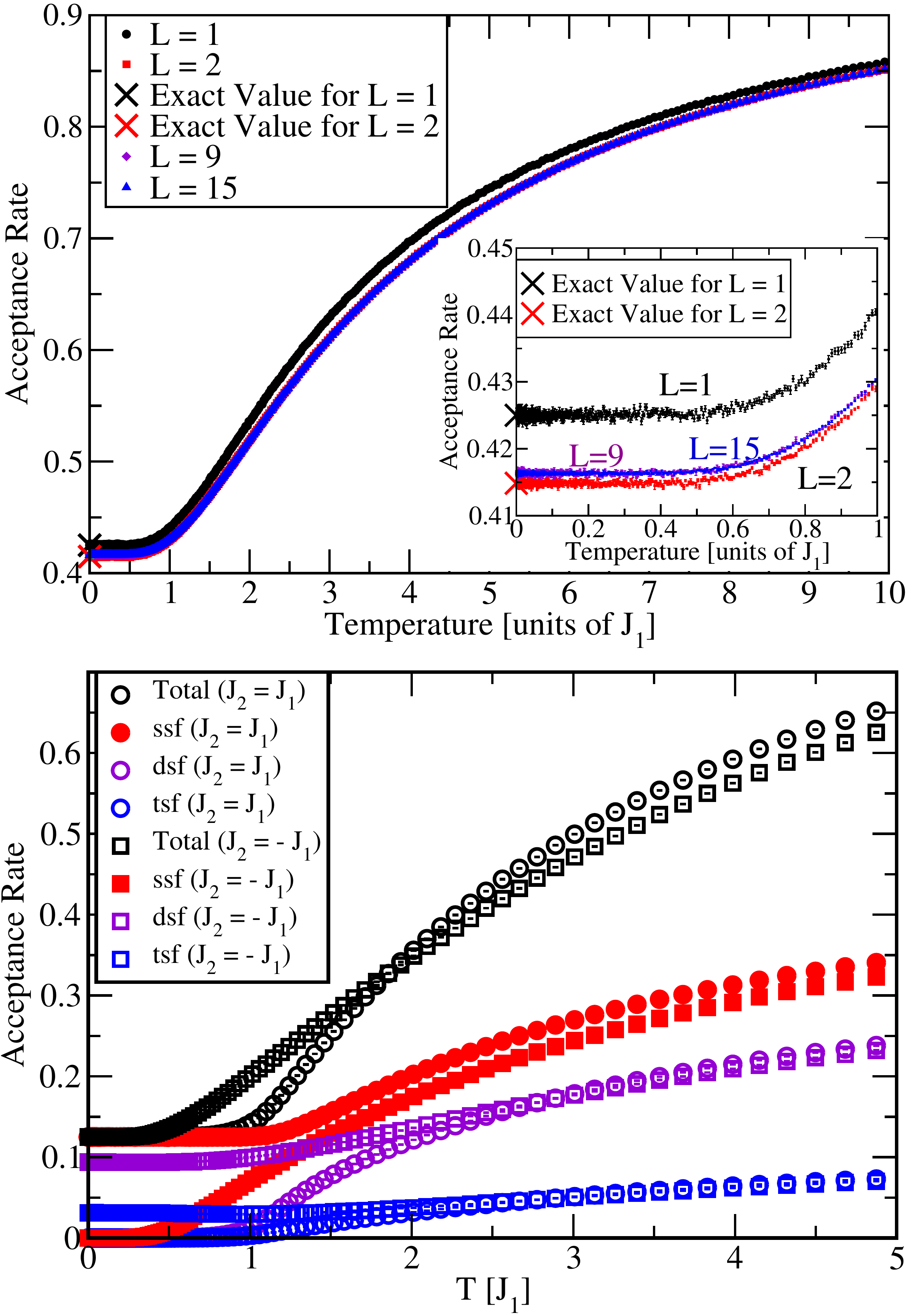}
 \caption{(Color online)(a) Acceptance rate versus temperature for $J_1=1$, $J_2=0$.  Representative lines for $L=1,2,9$ and $L=15$ are shown.  In the inset, we see that the zero temperature $L$=1 acceptance rate is considerably larger than the acceptance rate for $L=2$ and above.   Exact ground state results for $L=1,2$ calculated as discussed in the text agree with the $T\rightarrow 0$ limit of our simulations as shown by the horizontal lines in the inset and X's in the main figure. (b) Acceptance rate for $J_2\ne0$ ($L=4$) with hybrid spin flip processes allowable as discussed in Sect. \ref{sectionhsf}.  For $J_2>0$, as $T\rightarrow$0, twenty-five percent of all attempted single spin flip processes are accepted, corresponding to a magnetically ordered ground state with completely free spins at the 6-coordinated sites.  For $J_2<0$ single spin flip processes freeze out at low temperatures, while triple spin flip and double spin flip processes continue to allow the system access to all ground states at low temperatures.  \label{figure2} }
\end{figure}



\subsection{Acceptance rate}

\subsubsection{Single spin flip}

For the pure sorrel net ($J_2=0$) we have carried out single spin flip Monte Carlo simulations and find a non-negligible acceptance rate for all $L$ ($\approx 41.5\%$ for $L\ge 2$) even in the zero temperature limit as shown in Fig. \ref{figure2} (a).  For small system sizes ($L=1,2$), we can find exact results for the zero temperature acceptance rates (0.425 and 0.4148376 respectively).  These are shown in the inset to Fig. \ref{figure2} (a) to agree well with our simulations.  The $L=2$ acceptance rate appears to lie slightly lower than the acceptance rate for higher values of $L$, which appear to reach a constant value well before $L=15$.

When $J_2>0$ the single spin flip process continues to admit flips on the 6 coordinated sites to the lowest temperatures, approaching a $25\%$ acceptance rate.  As the degrees of freedom associated to the spins at the remaining sites freeze out (as one expects from the detailed nature of the ground state configurations shown below, see Fig. \ref{figure4}), the acceptance rate is not a monotonic function of temperature, dipping below 25\% prior to recovering as $T\rightarrow 0$.

When $J_2<0$, single spin flip simulations are insufficient to allow the system access to its degenerate ground state, and the acceptance rate dies quickly as $T\rightarrow 0$, indicating a lack of ergodicity as has been commonly found in the dipolar spin ice systems, where loop algorithms have commonly been introduced.\cite{loops} 

\subsubsection{Hybrid spin flip} \label{sectionhsf}

To address the concern that single spin flip processes alone may not allow our spin system to properly equilibrate for $J_2\ne 0$, given that in the ground state the spins joined by the $J_2$ bonds want to align(anti-align) when $J_2<(>)0$, we have developed a hybrid spin flip algorithm.   The addition of these flipping processes is in the spirit of loop algorithms\cite{loops} which have been deemed necessary in other frustrated magnetic systems to restore ergodicity to the system at low temperatures. To implement this hybrid spin flip algorithm, at each update attempt we randomly choose a site in the lattice.  If this site belongs to a $J_2$ bond, we randomly choose to either flip the spin (single spin flip (ssf)), or to flip the pair of $J_2$-bonded spins (double spin flip (dsf)).  If the chosen site belongs to a 6 coordinated site, then we randomly choose to either flip the spin (ssf) or to flip the spin and one of its randomly choosen nearest $J_2$ bonded pairs of neighboring spins (triple spin flip (tsf)).  Since 3/4 of all sites in the lattice belong to a $J_2$ pair, dsf processes are attempted on average during 3/8, tsf processes 1/8, and ssf processes 1/2 of the update attempts.  With this hybrid code, one sees (in Fig. \ref{figure2} (b)) that the total acceptance rate remains finite to zero temperature for all $J_2\ne 0$. 

\paragraph{$J_2 > 0$}

As was found using ssf for an antiferromagnetic bond between the $J_2$ bonded sites, at the lowest temperatures only 1/4 of the spins remain active, and 1/4 of the ssf attempts within the hybrid spin flip are accepted.  However, the presence of dsf and tsf processes does allow the temperature evolution of the ssf acceptance rate to become smooth and monotonic.  This indicates that while the dsf and tsf processes effectively turn off once the system has achieved a partially ordered ground state (as described below), their presence at higher temperatures allows the system to remain in equilibrium as this order occurs.

\paragraph{$J_2 < 0$}

At low temperatures with a ferromagnetic bond between the $J_2$ bonded sites, ssf processes are no longer accepted.  However tsf and dsf processes allow the system to access its degenerate ground states and move from one to another.

\paragraph{$J_2 = 0$}

To test the validity of this approach, we have run hybrid spin flip processes for $J_2 = 0$.  These give results which agree with those of the ssf code.

\begin{figure}
\includegraphics[scale=0.32]{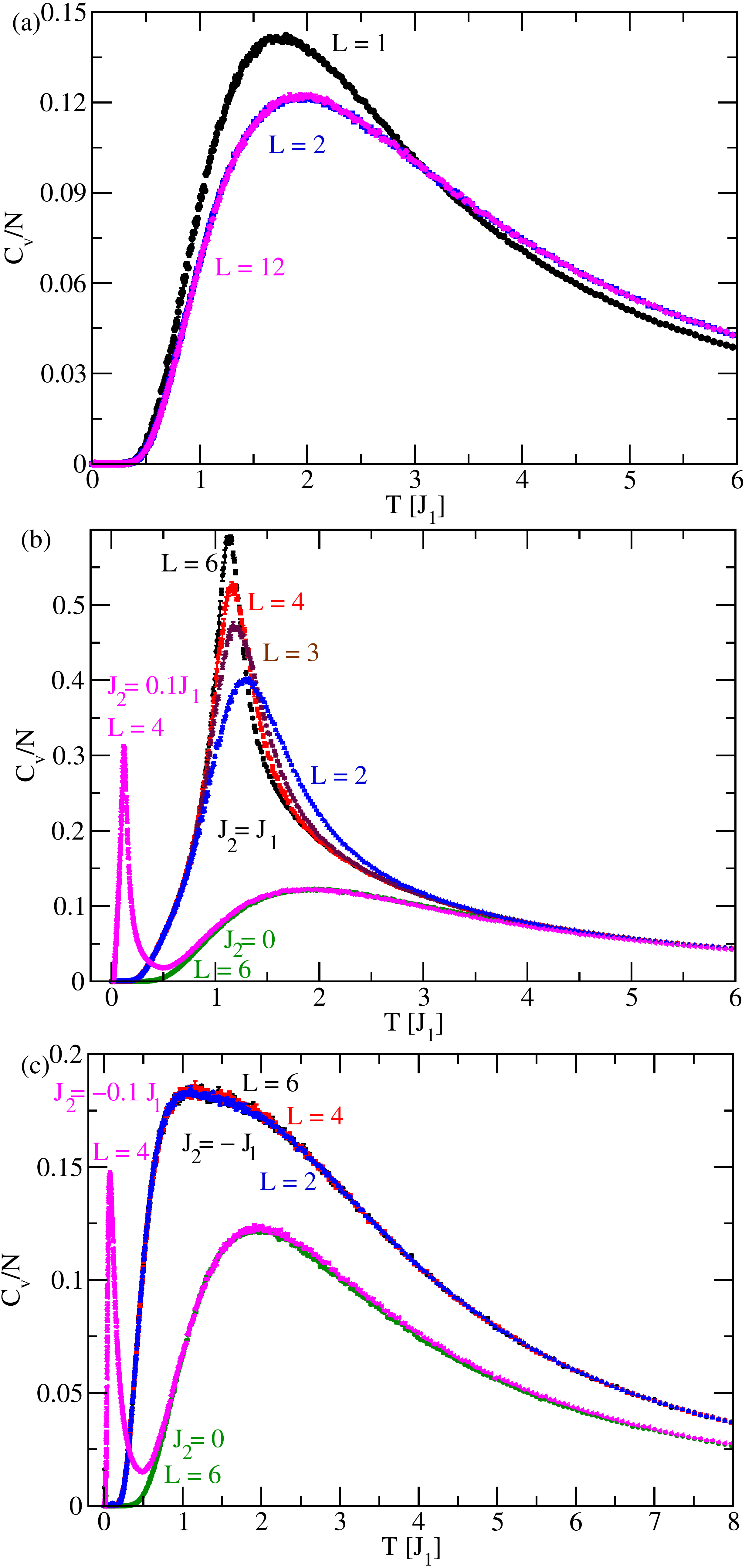}
\caption{(Color online)  (a) Heat capacity vs. temperature for $J_2$ =0.  For $L\ge 2$ heat capacities are very similar. For $J_2\ne0$ ((b) and (c)) a sharp peak is seen below the $J_2 = 0$ peak.  For $J_2 > 0$, (b), this peak sharpens with increasing system size.  For $J_2 < 0$, (c), both peaks show remarkably little variation with $L$.  
\label{figure3}}
\end{figure}

\section{Results}

\subsection{Heat capacity}

As shown in Fig. \ref{figure3} (a), the heat capacity,
\begin{equation}
C_V(T) = \frac{< \! E^2 \! > - < \! E \! >^2}{T^2}\text{,}\label{equation1}
\end{equation}
for the pure sorrel net shows a broad peak indicating the entrance of the system to a degenerate ground state which features two spins up(down) and one spin down(up) on each of the corner-shared triangles.  Here $E$ is the energy of a particular spin configuration on the lattice and $T$ is the temperature.   The heat capacity exhibits a mild dependence on system size, with the $L=1$ peak being noticeably larger than subsequent $L$.  There is little variation in the heat capacity for $L=2$ up to $L=18$.  In Fig. \ref{figure3} (a) we show heat capacity versus temperature curves for $L = 2$ and $L = 12$ to illustrate that finite size effects are likely small.

Turning on the coupling $J_2$, we see a sharp peak in the heat capacity well below this broad feature for both signs of $J_2$. As $J_2$ increases, this sharp feature moves to higher temperatures, broadens, and becomes less distinct from the peak representing the crossover to a two up (down) and one down (up) spin state.  In both cases, this secondary peak eliminates much of the entropy associated to the pure sorrel net at temperatures which increase with the magnitude of $J_2$.  As seen in Fig. \ref{figure3} (b), when $J_2 > 0$, as $L$ increases the second peak in the heat capacity sharpens as one might expect from a phase transition to a long range ordered state.  In strong contrast to this behavior, when $J_2 < 0$ as seen in Fig. \ref{figure3} (c), the second peak remains essentially unchanged as $L$ increases, indicating a crossover to a short-range ordered state.
 \begin{figure}
 \includegraphics[scale=0.1]{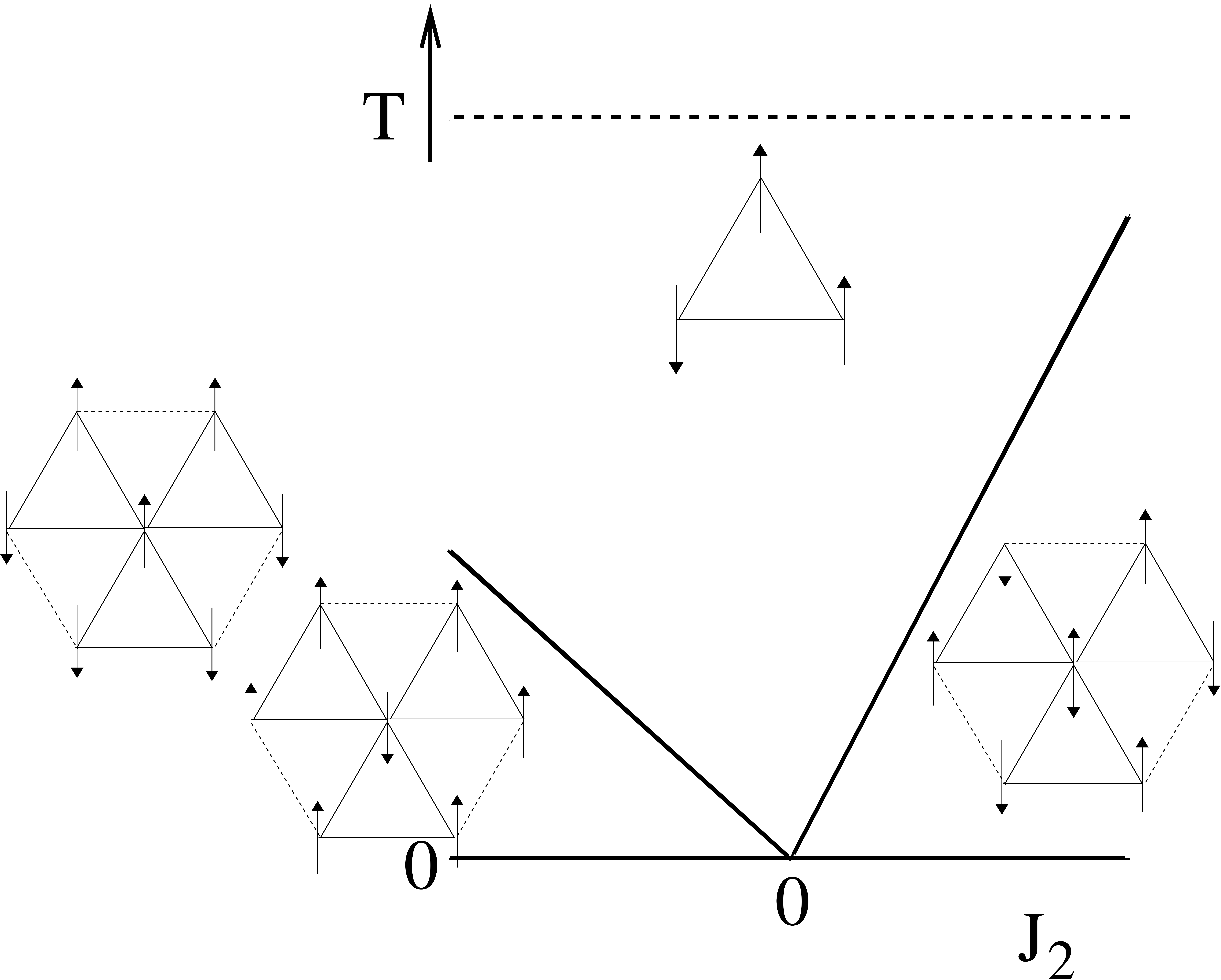}
 \caption{ A phase diagram illustrating the robustness of the sorrel lattice entropy to variations in $J_2$.  For $J_2>0$ at low temperatures one sees completely unconstrained spins on the 6 coordinated sites as illustrated.  For $J_2<0$ below the transition, satisfying the ferromagnetic $J_2$ bonds leaves the system with considerable remaining degeneracy.\label{figure4}}
 \end{figure}
 
\begin{figure}
\includegraphics[scale=0.35]{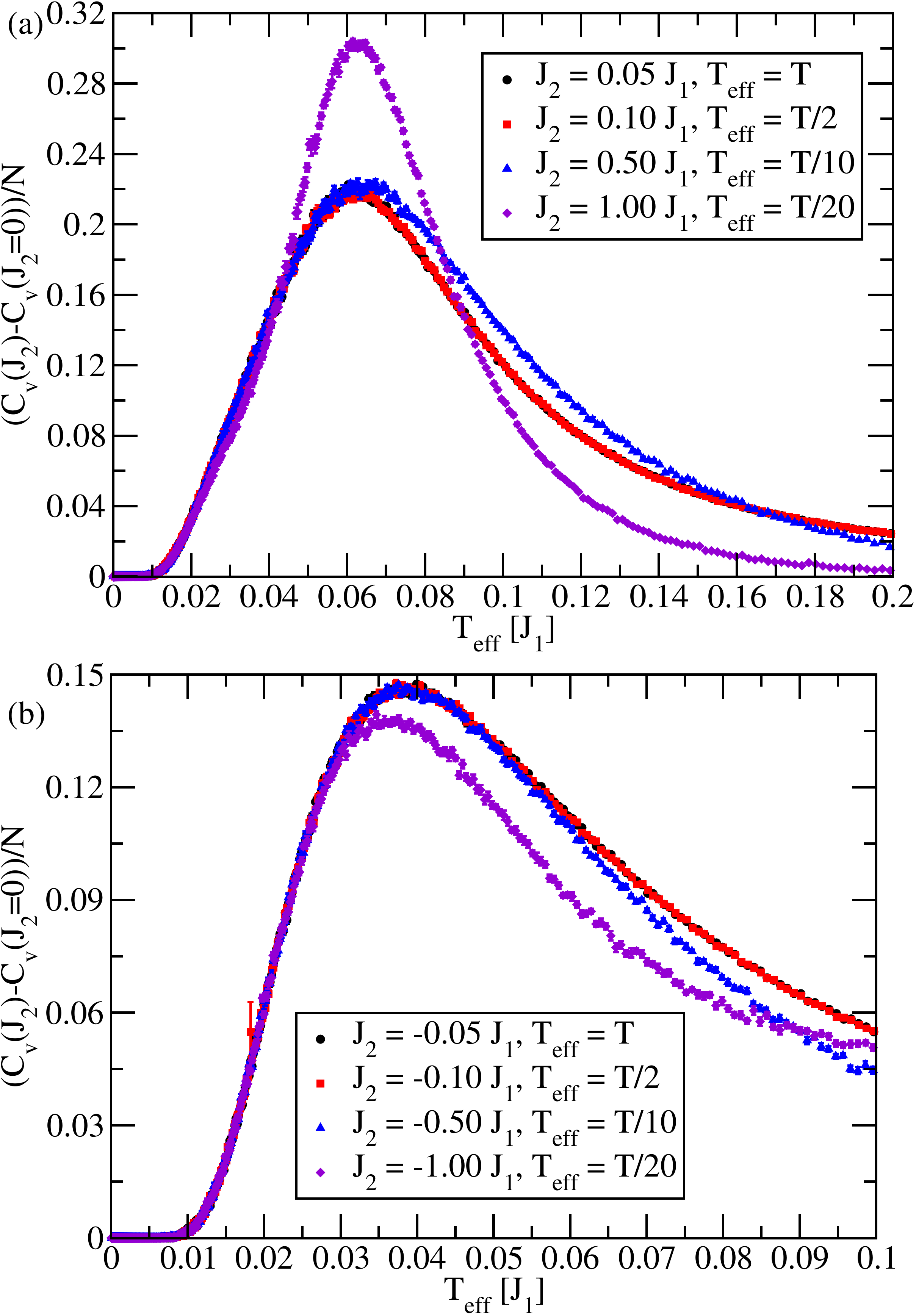}
\caption{(Color online) Heat capacity per spin at $J_2\ne 0$ minus the $J_2 = 0$ heat capacity per spin versus scaled temperature for $L = 2$.  (a) For $J_2 > 0$ the peak seems to occur at the same scaled temperature, indicating a linear dependence of the transition as shown in Fig. \ref{figure4}.  (b) For $J_2 < 0$ the crossover occurs at a lower temperature, and for small $J_2$ maintains a linear $J_2$ dependence at the critical temperature.\label{figure5}}
\end{figure}

\subsection{Phase diagram}

In Fig. \ref{figure4} we sketch the phase diagram for the Ising model on the generalized sorrel net.  In this plot, the dashed line represents the crossover from local disorder to each $J_1$ triangle having a net energy $-J_1$.  That is, below this line each triangle of the pure sorrel lattice adopts a spin structure with two spins up (down) and one spin down (up).  From the heat capacity, the temperature associated to this crossover corresponds to the maximum seen in the broad first peak of $C_v$ versus $T$, which appears not to change much as $J_2$ is turned on.  As the temperature is further decreased, for $J_2 > 0$ the spin system undergoes a transition to a long range partially ordered state, as pictured in Fig. \ref{figure4}.  In this state the 6-coordinated spin sites are completely free Ising spins, while all other nearest neighbor spin pairs feature antiparallel spins.

For $J_2 < 0$, as the temperature decreases there is a (size independent) crossover to a short-range ordered state, where across each $J_2$ bond the spins are parallel.  Every pair of parallel spins along $J_2$ bonds is possible, with the 6-coordinated spin site adopting a unique spin configuration antiparallel to the dominant spin configuration of its 3 nearest neighbor pairs.  One sees that such states are connected by the above-described dsf and tsf updates, and not by ssf updates.

At $J_2 = 0$, the spin system remains disordered to zero temperature classically.  As $J_2$ turns on, the peak corresponding to the transition/crossover to a long range ordered/short range ordered state occurs at proportionally larger temperatures, as indicated by the straight lines in our sketched phase diagram.  If the high temperature crossover and its associated temperature dependent heat capacity is truly independent of $J_2$, then by subtracting out the $J_2 = 0$ peak one would be left with the second peak alone.  Further, if one then scaled the temperature of several of these curves by their value of $J_2$, their peaks would be expected to coincide in the case that the transition temperatures scale linearly with $J_2$.  Such plots (for $L=2$) are shown in Fig. \ref{figure5} (a) (for $J_2 >0$) and (b) (for $J_2 < 0$). One sees that the temperature corresponding to the peaks of these curves does show a remarkable coincidence in both cases.  In fact, only the peak of $J_2 = -J_1$ appears to lie marginally below this linear temperature dependence. 




\subsection{Entropy}

  At very high temperature one expects there to be essentially no correlations between the Ising spins, or $2^N$ equally weighted spin configurations for an entropy per spin of $\ln(2)$.   As one lowers the temperature, the residual entropy in the system decreases.  In ordered materials below the ordering transition temperature the entropy quickly approaches zero.  In magnetically frustrated spin systems entropy can remain to very low temperatures.  As such it is interesting to investigate the behavior of the residual entropy in the system,
 \begin{equation}
 S(T) = S(T=\infty) - \int_T^\infty \! \frac{C_V(T)}{T} \, dT\text{,}\label{equation2}
 \end{equation}
and the zero temperature limit $S(T=0)$ of this quantity.

 \begin{figure}
 \includegraphics[scale=0.35]{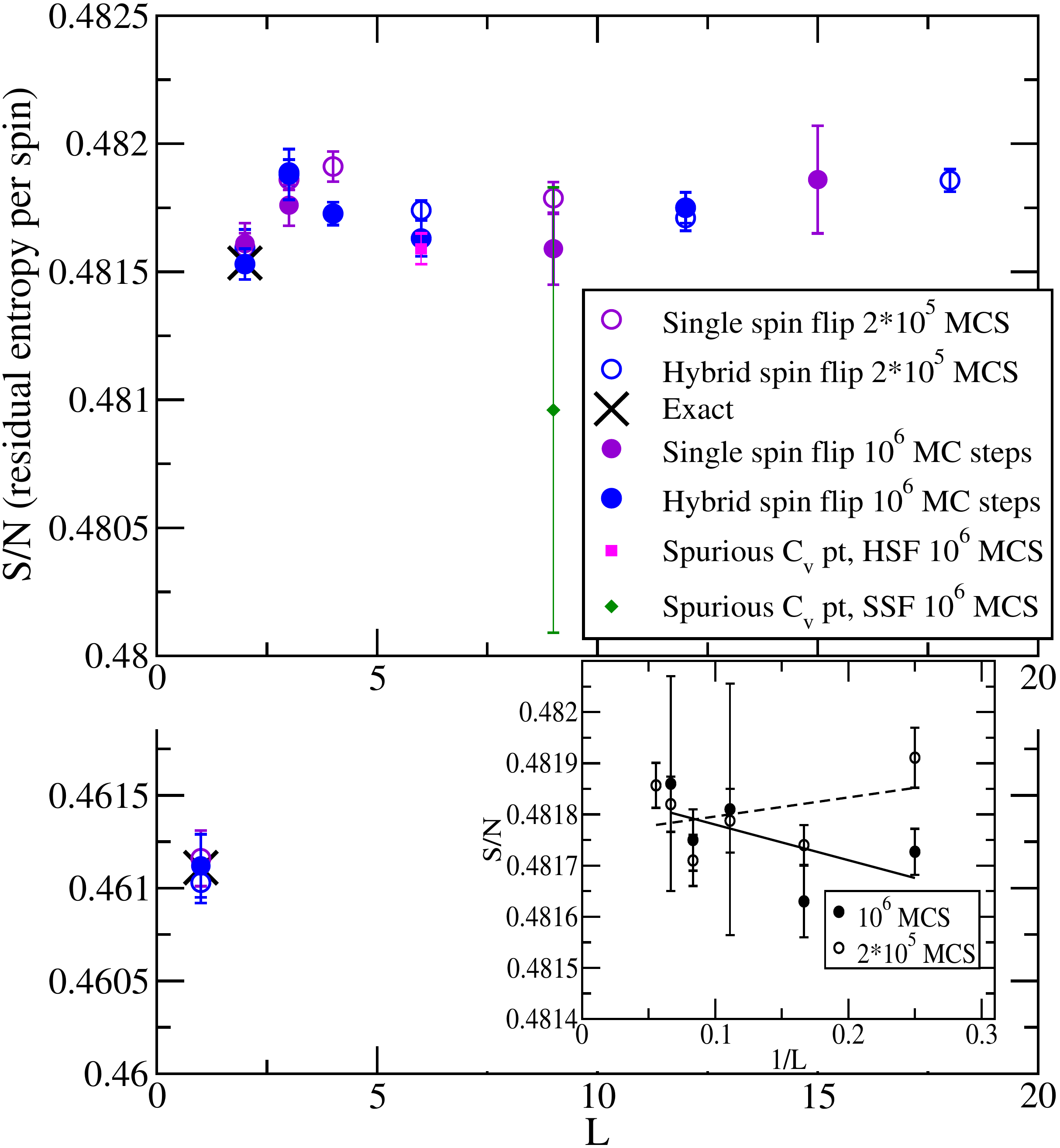}
 \caption{(Color online) Residual entropy (as $T \rightarrow 0$) versus system size ($L$) is roughly constant beyond $L = 2$.  Shown are data from two different equilibration times with the same initial seeds and the effect on the heat capacity of a single spurious low temperature point in the heat capacity.  Simulations agree well with exact results at $L = 2$ and $L = 1$, which is shown on a different scale.  (Inset) Residual entropy versus $\frac{1}{L}$ for $L\ge 4$ is used to extract a residual entropy $\frac{S}{N} = 0.48185\pm 0.00008$ in the thermodynamic limit.\label{figure6}}
 \end{figure}
\subsubsection{Pauling estimate}

When $J_2 = 0$, if one makes the assumption that spins beyond their nearest neighbor are uncorrelated, then each spin site has two degrees of freedom constrained by the fraction of the possible spin configurations on each triangle which belong to the ground state, which for a triangle-based lattice is $\frac{6}{8}$.  On the sorrel net each spin belongs to either two or three triangles, so that there are $N_t = \frac{3N}{4}$ triangles in total.  The Pauling degeneracy of the ground state  is then $ 2^N(\frac{6}{8})^{N_t}=2^N(\frac{6}{8})^{ \frac{3N}{4}}$, hence the Pauling estimate for the entropy per spin is $\frac{S}{N} =\ln(2(\frac{3}{4})^{\frac{3}{4}})=\frac{1}{4}\ln(\frac{27}{4})\approx 0.477386$.
 \begin{figure}
 \includegraphics[scale=0.35]{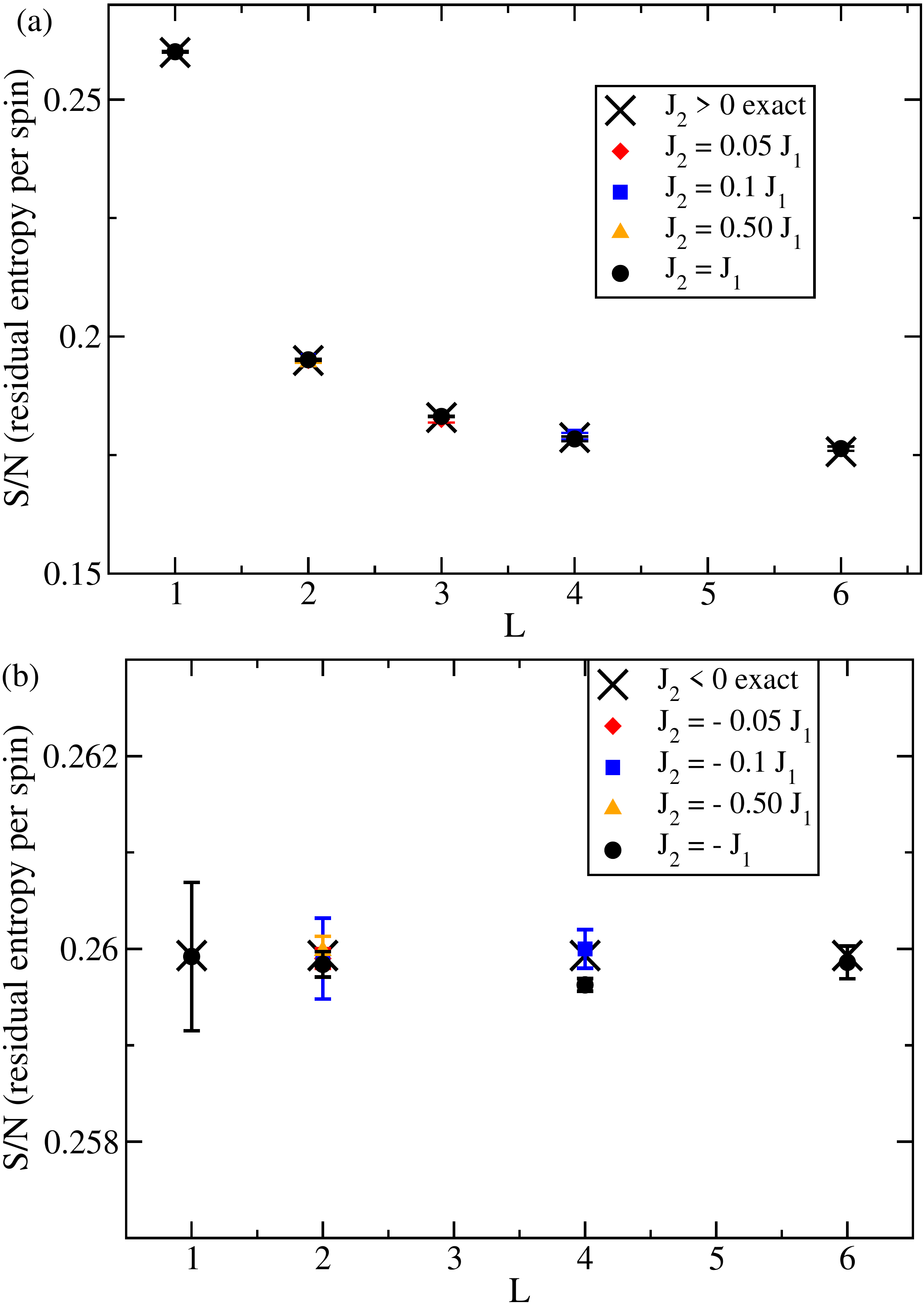}
 \caption{(Color online) As discussed in the text, we have found exact results for the ground state residual entropy for $J_2 \ne 0$.  These agree well with MC simulations for (a) $J_2 > 0$ and (b) $J_2 < 0$.\label{figure7}}
 \end{figure}
\subsubsection{Exact}
When the number of spins is small, it is numerically possible to count the number of ground states that satisfy the ground state conditions for $J_2=0$.  For any spin configuration in the ground state two of the spins on any given triangle must be up(down) while the remaining spin is down(up).  So doing, for L=1 we find the ground state to have 40 distinct members while for L=2 it has 4921350 distinct states.  This indicates that the exact result for the residual entropy of the pure sorrel net is $\frac{S}{N} = 0.46111$ for L=1 and $\frac{S}{N} = 0.481534$ for L=2, both numbers which fall within our error bars for the residual entropy as seen in Fig. \ref{figure6}.

Adding $J_2\neq 0$, the available ground states additionally must satisfy parallel(antiparallel) spins across the $J_2$ bond when $J_2<(>)0$. This means that each ground state is a subset of the sorrel lattice states.  For $L=1$ this reduces the number of degenerate ground states to just 8 independent of the sign of $J_2$.  For $L=2$ when $J_2>0$ only 512 of the ground states satisfy the condition, while for $J_2<0$ there are 4096 ground states.   We see that these numerically extracted exact values are the $L = 1$ and $L = 2$ limits of general exact formulae for the $J_2 \ne 0$ entropy.  For $J_2>0$ all the ground states feature a completely free spin on the 6 coordinated sites, and only one of two global spin configurations for the remaining spin sites, indicating that we have $2^{2L^2+1}$ degenerate ground states, and that the entropy per spin should therefore weakly depend on $L$: $\frac{S}{N} = \frac{1}{8L^2}\ln(2^{2L^2+1})=(\frac{1}{4}+\frac{1}{8L^2})\ln(2)$.   For $J_2<0$, the ground states feature parallel spins on each of the $J_2$ bonds, meaning that each 6-coordinated site has a definite orientation antiparallel to the dominant spin type of its three neighboring $J_2$ bonded pairs of spins.  However, every combination of $J_2$ bond parallel spins is possible, so there are $2^{3L^2}$ ground states, for a ground state entropy per spin of $\frac{S}{N} = \frac{3\ln(2)}{8}$.  Hence for all $J_2 \ne 0$ we have exact results for the residual entropy.

\begin{figure}
\includegraphics[scale=0.35]{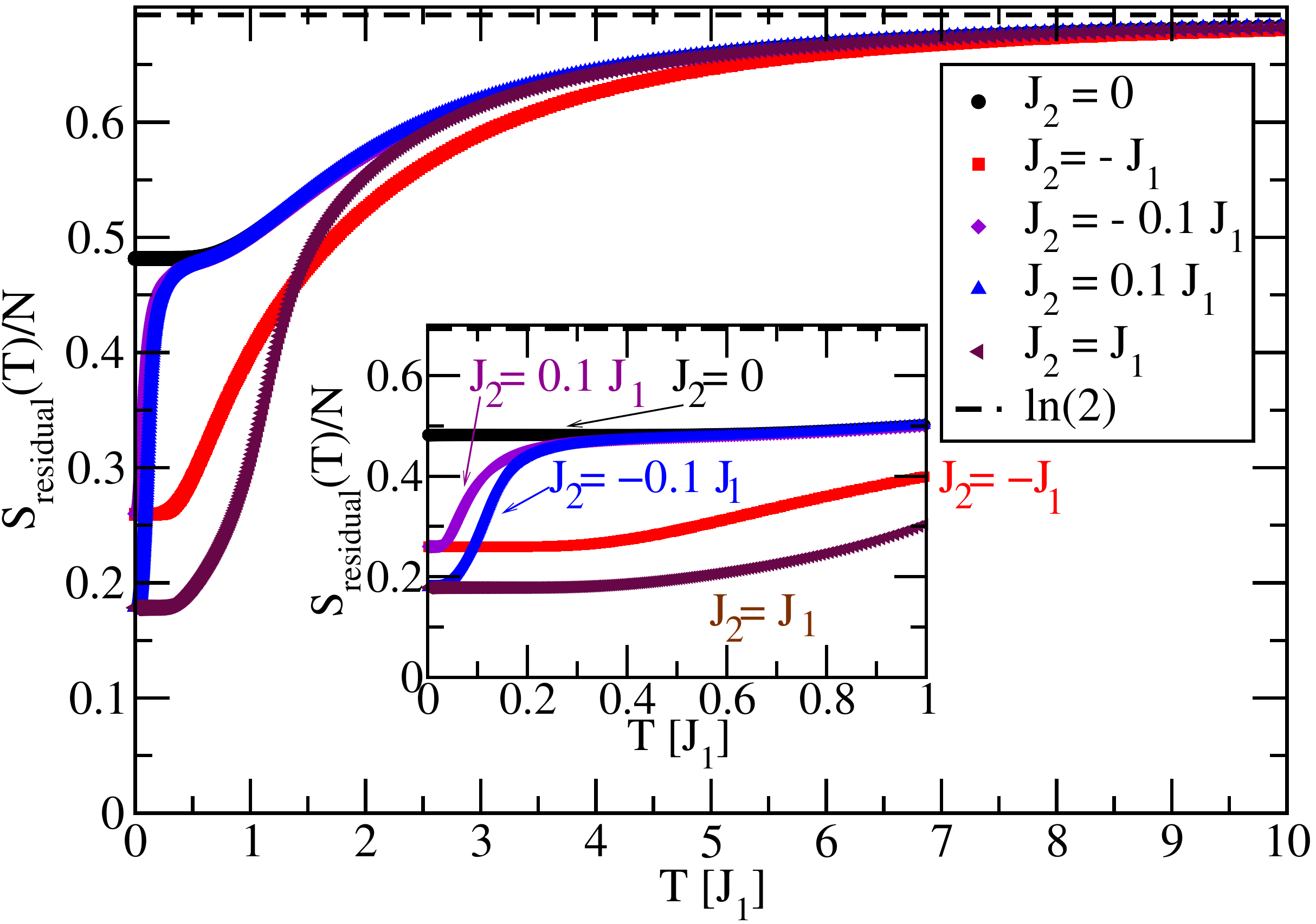}
\caption{(Color online) The entropy remaining in the system per spin as a function of temperature for $L = 4$.  At high temperatures, all curves asymptotically approach $\ln(2)$.  For $J_2 = 0$, the residual entropy smoothly decreases until reaching a constant low temperature value $\frac{S}{N} = 0.48173\pm 0.00005$.  For small $J_2$, the entropy remaining follows the $J_2 = 0$ curve to quite low temperatures prior to sharply downturning to the exact $T = 0$ residual entropies shown in the text.\label{figure8}}
\end{figure}

\subsubsection{Monte Carlo}

As shown in Fig. \ref{figure6}, the residual entropy per spin calculated by Monte Carlo simulations\cite{note3} agrees well with the exact results at $L=1$ and $L=2$ for the pure sorrel net.  Fig. \ref{figure6} additionally presents data using both the ssf and hsf codes.  We see that simulations from both codes agree with the exact results and give equally valid results at larger $L$.  We further note that for $L=2$ and beyond, the residual entropy per spin for this case ($J_2 = 0$) appears to have reached a constant value (0.48185$\pm$0.00008)\cite{fits} as a function of system size, indicating that the thermodynamic limit has been reached well before our maximum system size of $L=18$, which corresponds to 2592 spins.   

As expected from our exact results, much of this residual entropy is quenched at very low temperatures when $J_2\neq 0$.  Our simulations indicate that the residual entropy per spin for $J_2<0$ remains constant as a function of $L$. All calculated residual entropies for $J_2 < 0$ agree within error of the exact results, except for the case $J_2 = -J_1$ and $L=4$, where a marginally lower entropy is found.  The reason for this small discrepancy has not been discovered as yet.  To calculate the errors for these results we numerically integrated four (or more at small $L$) heat capacities divided by temperature to find entropies versus temperature. We then found the average entropy at each temperature and its standard deviation.  As such, our stated errors may be a little small as they do not explicitly account for errors in the numerical integration procedure.  

 For $J_2>0$ the residual entropy per spin is consistent with the exact results with all cases lying within two standard deviations.\cite{except}

In Fig. \ref{figure8} we plot the entropy versus temperature for various values of $J_2$ at $L=4$.  We see that at high temperatures as $J_2$ is turned on from zero, all the curves follow the $J_2=0$ entropy.  As the temperature continues to decrease, the larger magnitude $J_2$ simulations begin to substantially deviate from the $J_2 = 0$ entropy in a manner consistent with the phase diagram as extracted from the heat capacity in Fig. \ref{figure4}.  When $|J_2|$ is small, this deviation does not occur until quite low temperatures, and when it does occur, the transition is quite sharp to a state with the ground state entropy. If a system can be made for which the Ising model on the sorrel lattice with $J_2 << J_1$ is a good approximation, one would expect to see a plateau in the residual entropy of the pure sorrel lattice over a certain temperature range, as shown in the inset to Fig. \ref{figure8}.
   

\begin{figure}
\includegraphics[scale=0.42]{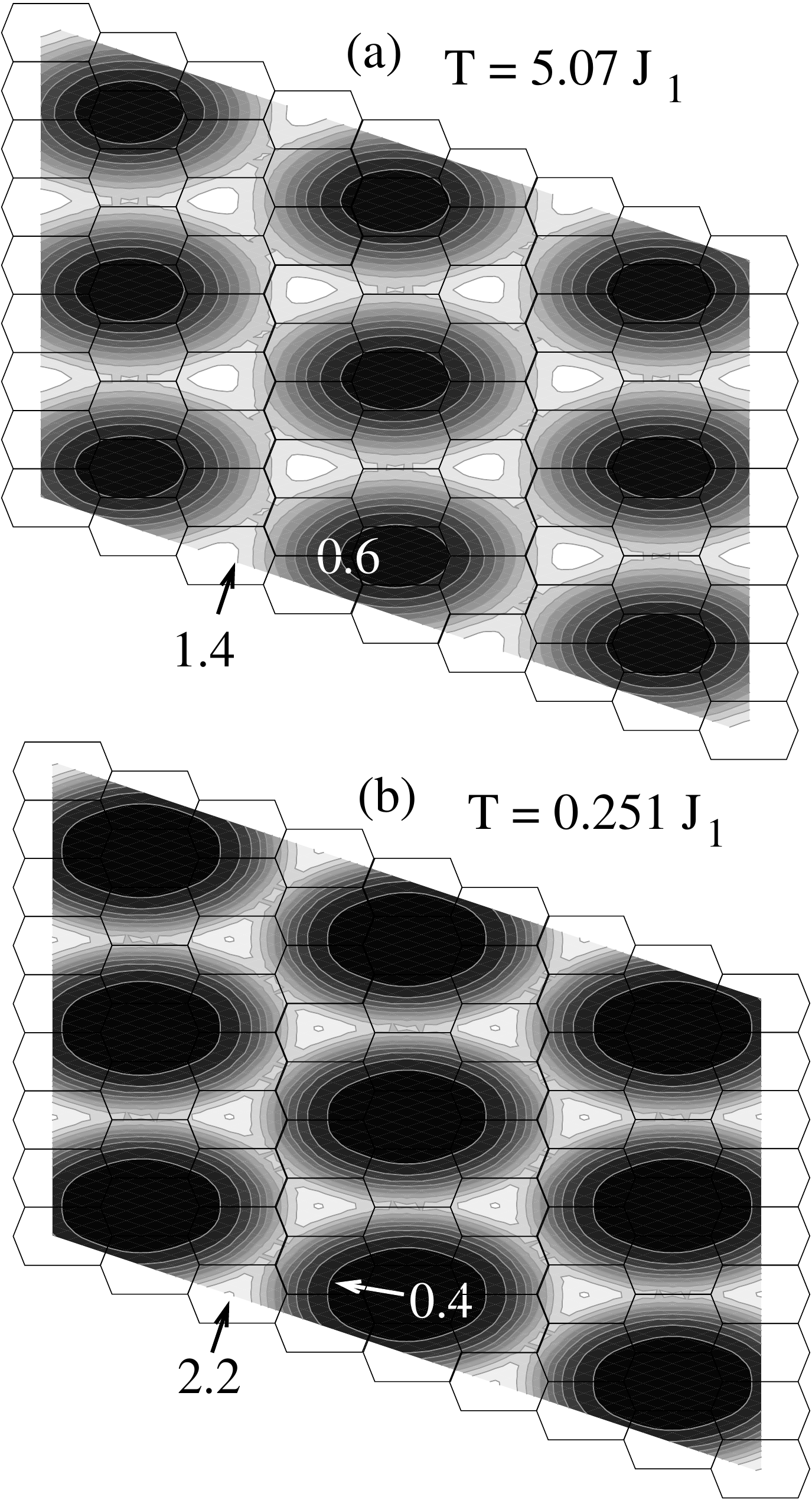}
\caption{For $L = 6$, an evaluation of the static structure factor at the indicated temperatures in reciprocal space for $J_2 = 0$, contour increments of 0.1 for (a) and 0.2 for (b).  The squished hexagon at the center represents the first Brillouin zone, which has been repeated to tile the reciprocal space.  The center of each hexagon indicates the number of reciprocal lattice vector steps to the centre from each side, the top center being (0,$\frac{16\pi}{\sqrt{3}a}$).  One sees diffuse weight about $\vec{q} = (\frac{4\pi}{a},0)$ and the 5 equivalent $C_6$ rotations about the origin at all temperatures.\label{figure9}}
\end{figure}

 \subsection{Spin-spin correlations}
%
%

The static structure factor provides a measure of the strength of the spin-spin correlations and their momentum dependence.  Where large single crystals of a material become available, the structure factor (convolved with an atomic form factor) allows a direct comparison between neutron scattering measurements and theoretical predictions.  Materials which order magnetically show the development of sharp, resolution-limited magnetic Bragg peaks at low temperatures.  In contrast, many frustrated magnetic systems show evidence of short range order in the form of disperse magnetic peaks concentrated around certain areas of reciprocal space.  Powder neutron scattering measurements average the angular dependence of such graphs, showing weight at finite wavenumbers.

\begin{figure}
\includegraphics[scale=0.35]{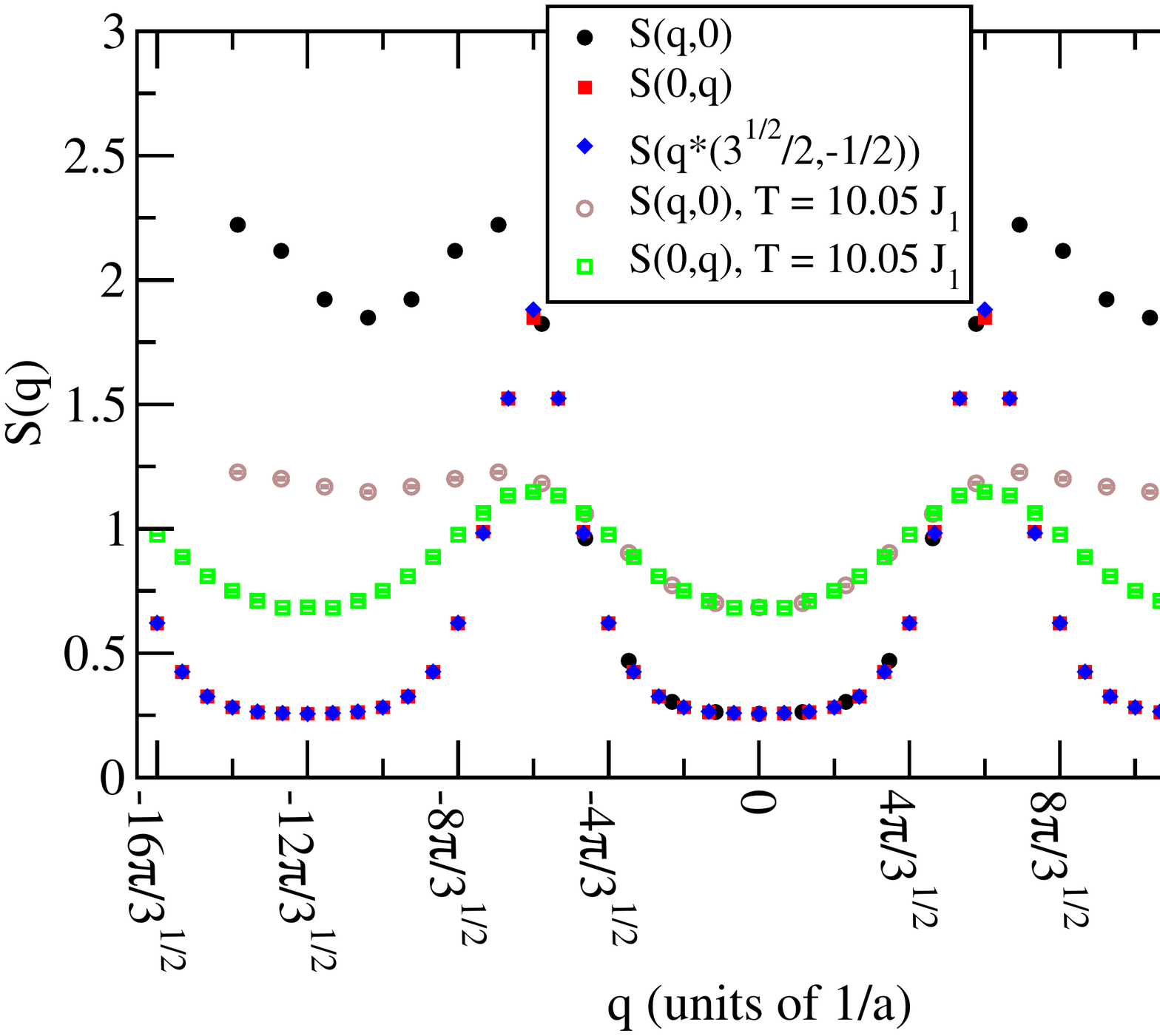}
\caption{(Color online) The ($L = 6$) static structure factor for $J_2 = 0$ at $T = 0.251 J_1$ and 10.05 $J_1$ along selected high symmetry directions.  Note that at both temperatures the weight remains symmetric about the origin to about $q = \frac{6\pi}{\sqrt{3}a}$.  Along $(0,q)$ the structure factor then repeats (note that this distance is 1.5 reciprocal lattice vectors).  Diffuse maxima occur about the points $\vec{q} = (\pm \frac{4\pi}{a},0)$ which are considerably less distinct at higher temperatures.\label{figure10}}
\end{figure}

Within Monte Carlo, the static structure factor is defined as,
\begin{equation}
{\mathcal{S}}(\vec{q}, T) = \frac{1}{N} \displaystyle\sum_{i,j} \! < \! \vec{s_i}(T) \cdot \vec{s_j}(T) \! > e^{i \vec{q} \cdot (\vec{r_i} - \vec{r_j})}\text{,}\label{equation3}
\end{equation}
where $N$ is the number of sites in the lattice $N=8L^2$, the summation runs over all pairs of spins on the lattice, the braces indicate the average over several MCS (20001 here) of the dot product between spins at sites $i$ and $j$, and $\vec{r_i} - \vec{r_j}$ is the vector between spin locations at sites $i$ and $j$.  For the sorrel lattice we have a hexagonal lattice with an eight site basis and reciprocal lattice vectors $\vec{b_1} = \frac{2\pi}{a}\left(1,-\frac{1}{\sqrt{3}}\right)$ and $\vec{b_2} = \frac{4\pi}{\sqrt{3} a}(0,1)$, where $a$ is the direct space lattice constant.  For a finite $8\times L\times L$ lattice, we can evaluate the structure factor at $\vec{q} = \frac{I}{L}\vec{b_1} + \frac{J}{L}\vec{b_2}$, where $I$ and $J$ are integers.  Taking $I$ and $J$ to run independently from $\{-4 L, 4 L\}$ allows us to gain a picture of the magnetic correlations well beyond the first Brillouin zone (shown as the central squished hexagon in Fig. \ref{figure9}).

\begin{figure}
\includegraphics[scale=0.42]{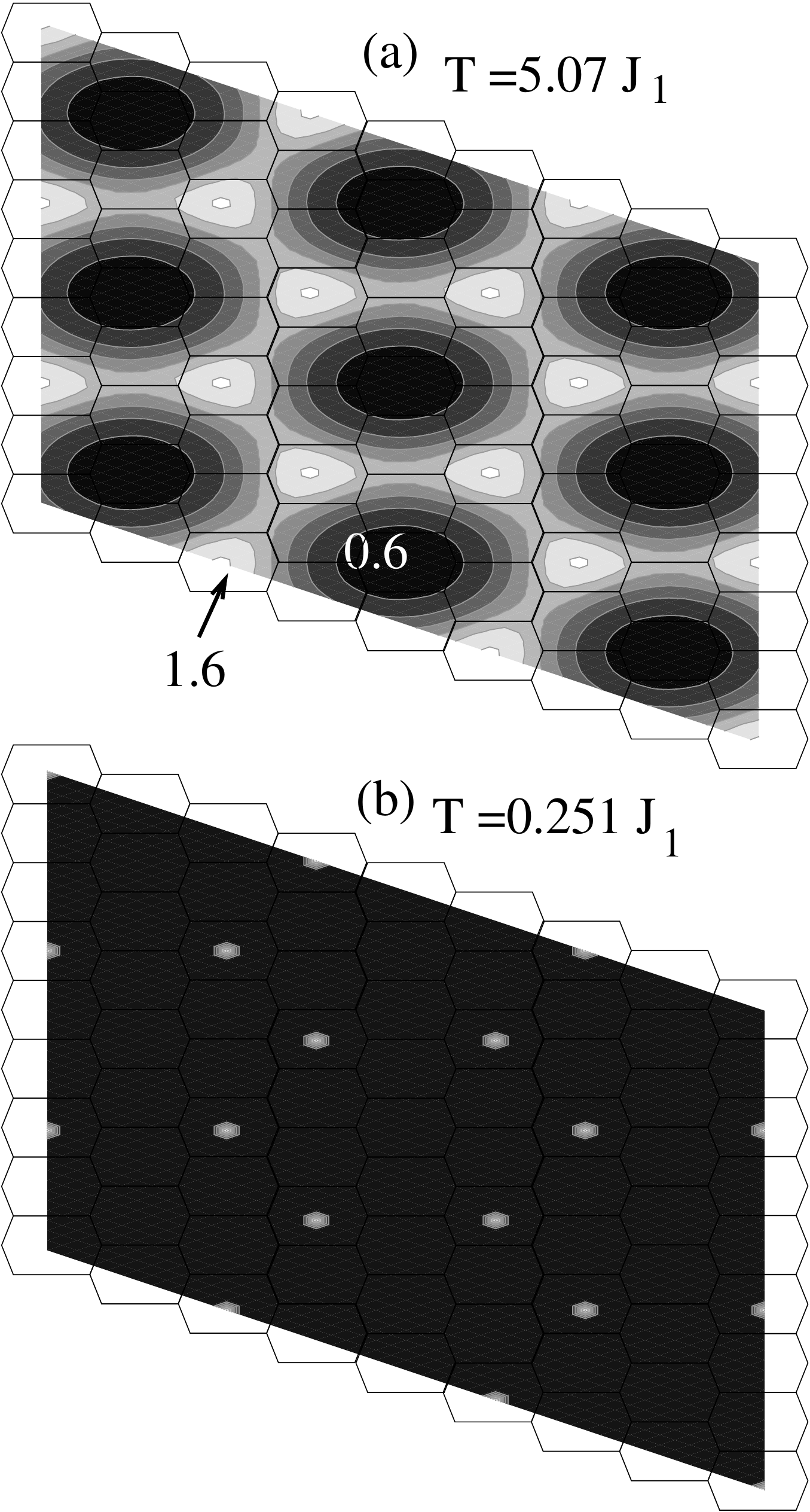}
\caption{For $L=6$ an evaluation of the static structure factor at the indicated temperatures for $J_2=J_1$ in reciprocal space. Contour increments of 0.2 in (a)  precede the resolution-limited Bragg peaks of (b).  The center of each hexagon indicates the number of reciprocal lattice vector steps to any point.  It is four reciprocal lattice vectors to the center from each side, the top center being $(0,\frac{16\pi}{\sqrt{3}a})$.\label{figure11}}
\end{figure}

\subsubsection{$J_2$ = 0}
When $J_2 = 0$, the structure factor weight is lowest in the vicinity of $\vec{q} = 0$, and isotropic at fixed $|q|$ as seen in Fig. \ref{figure9} and Fig. \ref{figure10}.  A disperse maximum is seen along the $(q,0)$ direction at $q = \pm\frac{4\pi}{a}$ and the equivalent $q = \pm\frac{8\pi}{a}$ and symmetry equivalent ($C_6$ rotation) points.  This corresponds to a linear combination of reciprocal lattice vectors, for instance, 2$\vec{b_1}$ + $\vec{b_2}$.  As the temperature is lowered, the relative intensity of this peak increases but the weight remains diffusely distributed around this point.

\begin{figure}
\includegraphics[scale=0.35]{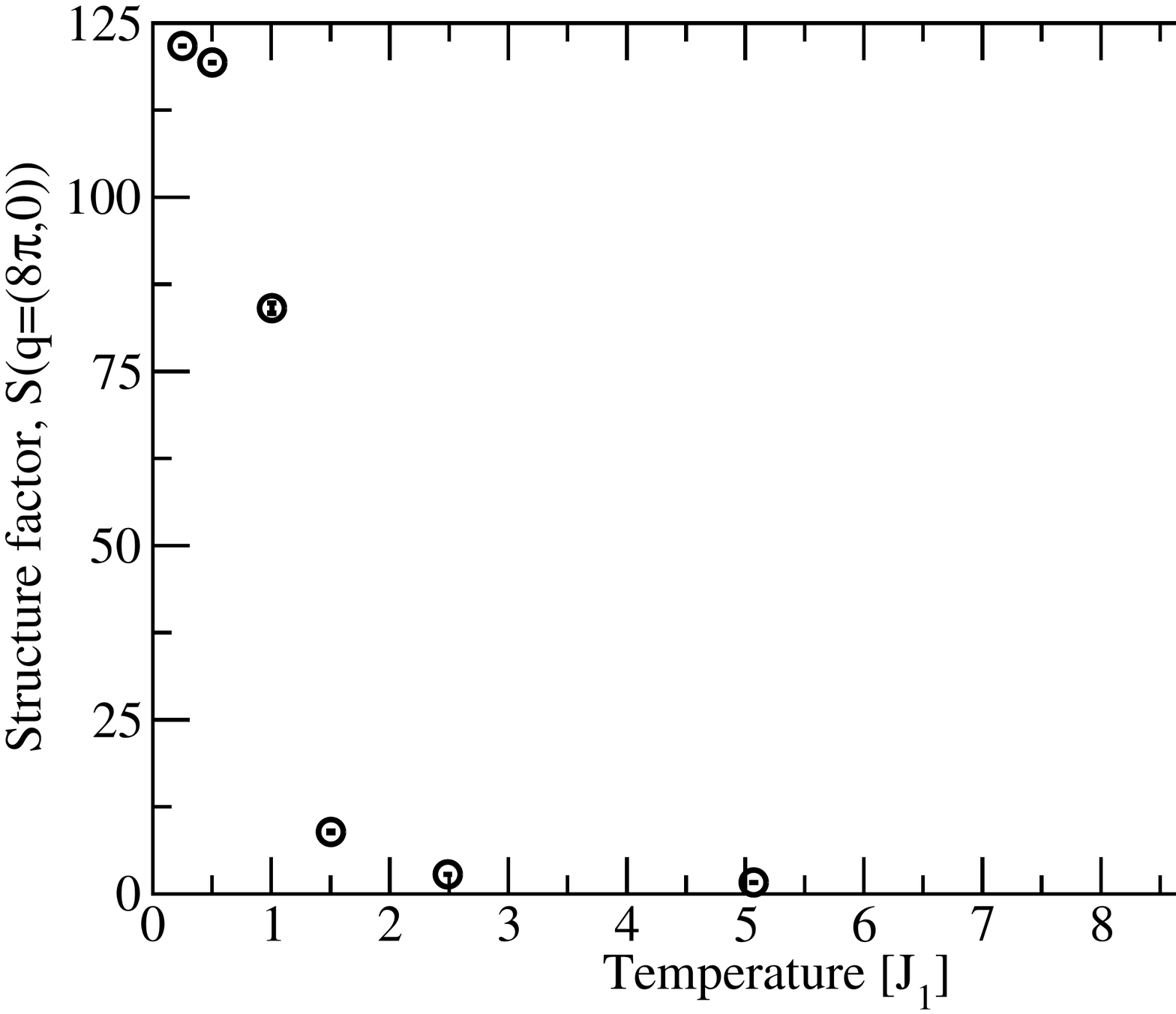}
\caption{\label{figure12} For $J_2 = J_1$ the temperature dependence of the peak of the structure factor at ${\vec{q}}=(\frac{4\pi}{a},0)$ and equivalent points.  By $T = 0.251 J_1$ it has almost reached the ideal ordered value as discussed in the text.}
\end{figure}

\subsubsection{$J_2 > 0$}
For $J_2\ne 0$ at high temperatures the structure factor appears quite similar to the $J_2 = 0$ case, as seen in Fig. \ref{figure11} (a) and Fig. \ref{figure13} (a).  However, for $J_2 > 0$, as the temperature is lowered one sees the rapid shifting of weight to the maximal direction--which becomes an ordering wavevector as shown in Fig. \ref{figure11} (b).  In Fig. \ref{figure12} we present the temperature dependence of the structure factor at this wavevector.  One sees that it saturates at a value of approximately 121.70 by the temperature $T = 0.25 J_1$.  In this partially ordered state we have argued that the spins at the 6-coordinated sites are completely free to fluctuate while the remaining spins form a magnetically ordered state.  If one simply assumes a spin structure with no moments on the 6-coordinate sites and antiferromagnetically ordered classical spins on the remaining sites, one finds a structure factor of 121.50.  It is interesting to note that this is not the full moment $\frac{3}{4}\cdot 288 = 216$ corresponding to an alignment along a particular axis of the spin structure, as one is used to from both ordered ferromagnets and antiferromagnets.  In the partially ordered antiferromagnetic state although it is a two-sublattice structure, one sublattice gains a weighting of $\frac{1}{2}$ on its moments due to the geometry of the lattice in its structure factor, lowering the total moment.  Adding back the self correlation of free (unit magnitude) spins gains an additional 0.25 contribution to this idealized structure factor, suggesting that by $T=0.25J_1$ the ordering state may have nearly reached the ideal partially ordered structure.

\subsubsection{$J_2 < 0$}

\begin{figure}
\includegraphics[scale=0.42]{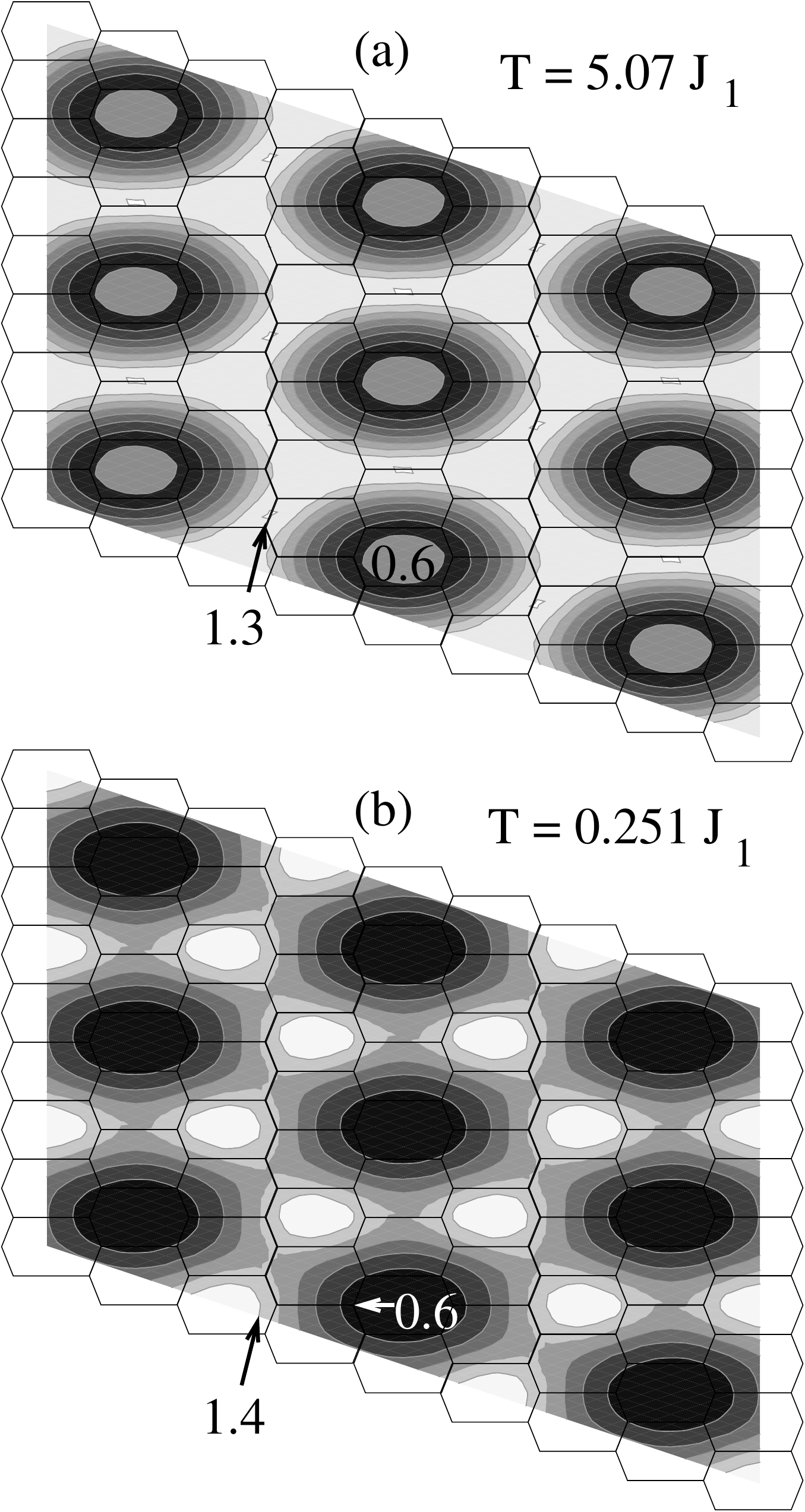}
\caption{For $L=6$ an evaluation of the static structure factor at indicated temperatures for $J_2 = - J_1$ in reciprocal space, contour increments of 0.1 (a) and 0.2  (b). The center of each hexagon indicates the number of reciprocal lattice vector steps to any point.  It is four reciprocal lattice vectors to the center from each side, the top center being $(0,\frac{16\pi}{\sqrt{3}a})$.  (a) At $T = 5.07 J_1$, the highest weight lies at $(0,\pm \frac{6\pi}{\sqrt{3}a})$ and equivalent positions, but considerable weight lies at $\vec{q} = (\pm\frac{4\pi}{a},0)$ which grows at low $T$ to form diffuse peaks.\label{figure13}}
\end{figure}

\begin{figure}
\includegraphics[scale=0.35]{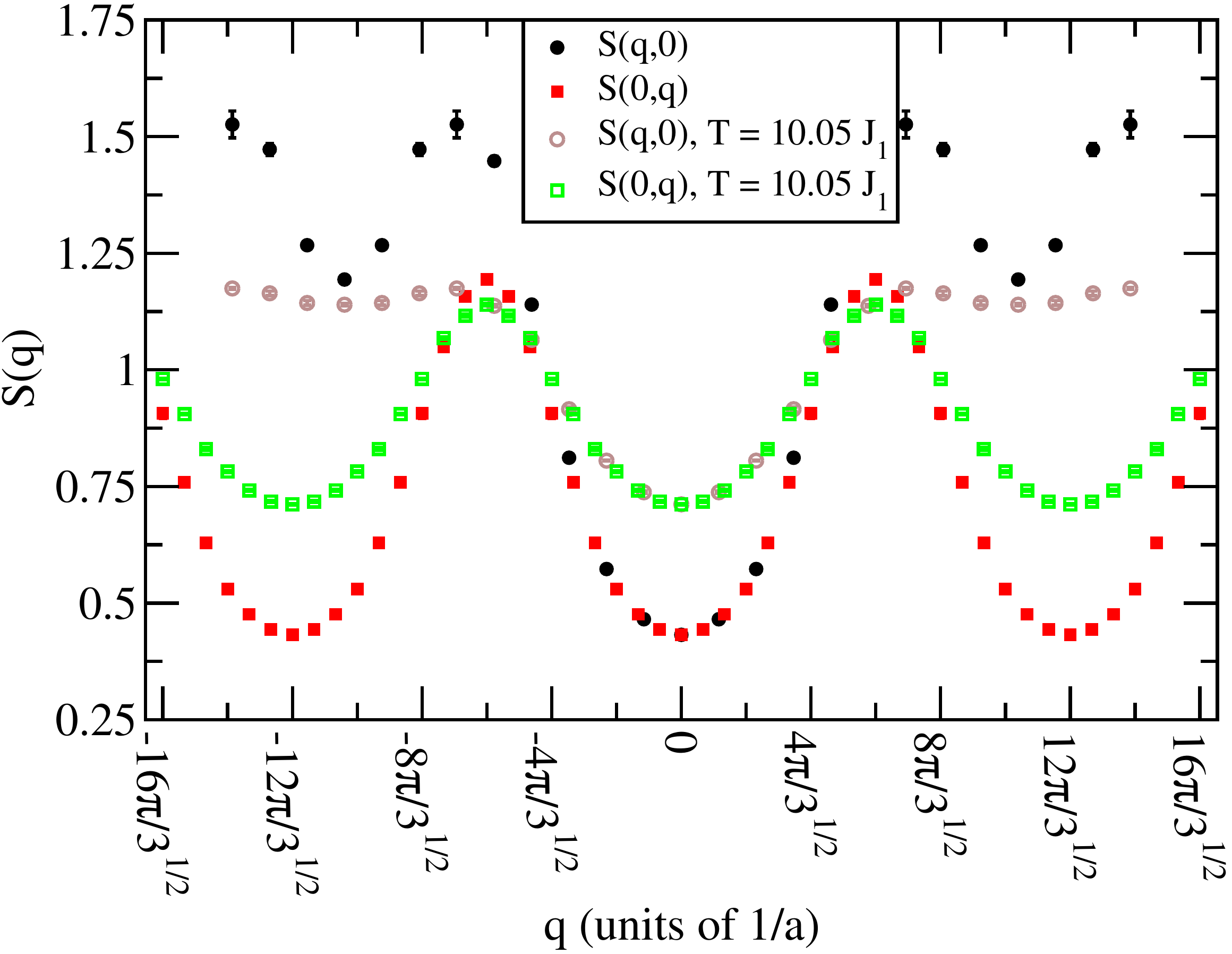}
\caption{The structure factor at low and high temperatures is now roughly symmetric only from $-\frac{4\pi}{\sqrt{3}a}$ to $\frac{4\pi}{\sqrt{3}a}$ and is less rounded than at $J_2 = 0$ (Fig. \ref{figure10}).  A diffuse maxima occurs at $\vec{q} = (\pm \frac{4\pi}{a},0)$ at low temperatures but washes out at higher temperatures. \label{figure14}}
\end{figure}

When $J_2 < 0$, the low temperature spin correlations (Fig. \ref{figure13} (b)) show once more (like when $J_2 = 0$) a broad feature about $\vec{q} = (\pm 4\pi, 0)$, $\vec{q} = (\pm 8\pi, 0)$ and 6-fold symmetry related points.  This feature sharpens as the temperature lowers, as can be seen in Fig. \ref{figure14}. Additionally, the contours appear less circular, more hexagonal than the $J_2 = 0$ case.

\subsection{Magnetic susceptibility}
\begin{figure}
\includegraphics[scale=0.30]{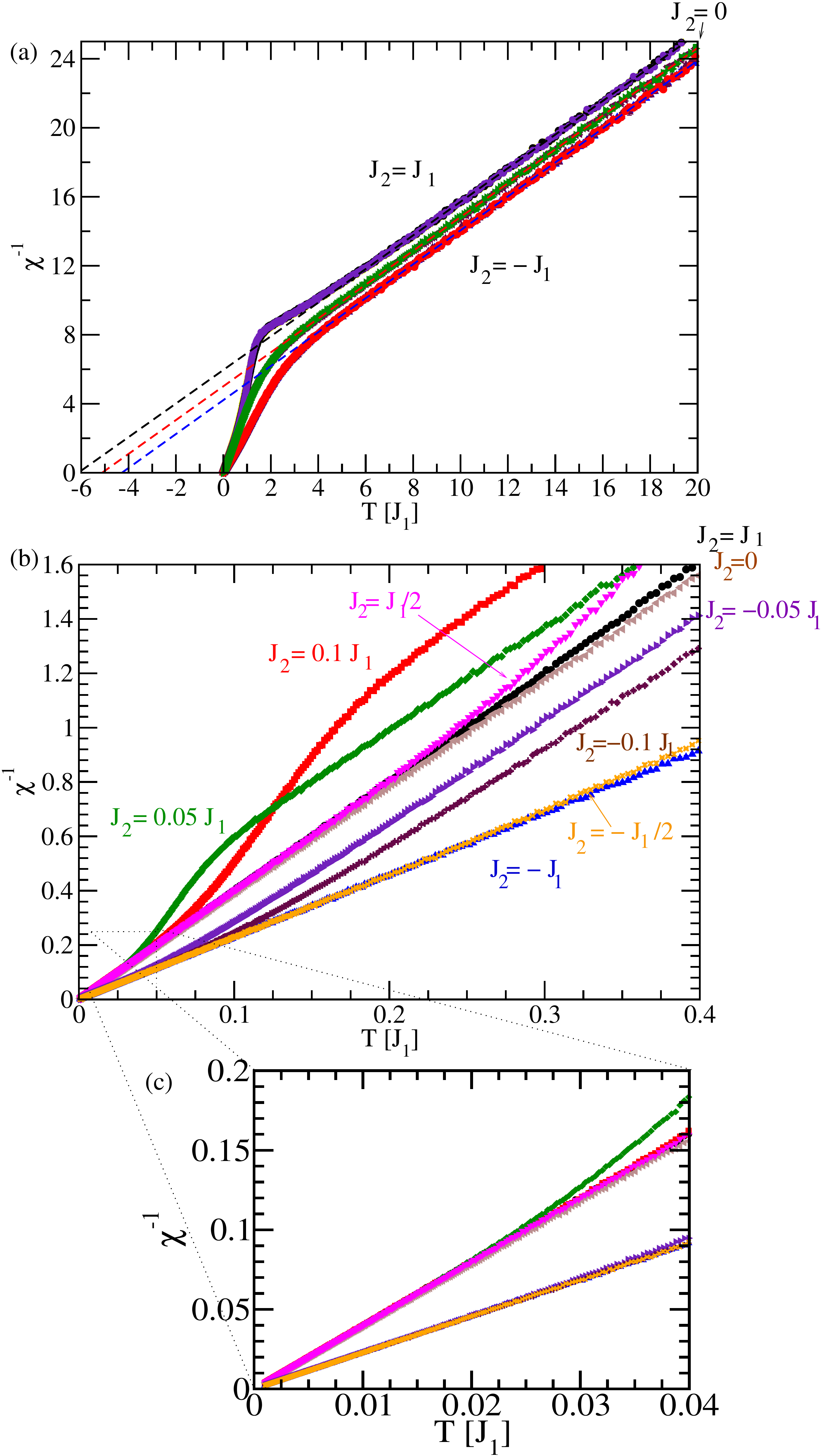}
\caption{(Color online) Inverse magnetic susceptibility versus temperature. (a) A high temperature Curie-Weiss intercept gets increasingly negative as $J_2$ increases.  Shown are $L =2$ and $L=4$ for all curves, and additionally $L=6$ for $J_2 = -J_1$.  There is very little system size-dependence of any part of the susceptibility beyond $L=2$.  (b) Variation of $J_2$ for $L=2$.  Curves asymptote at low temperatures towards different Curie laws depending on the sign of $J_2$, approaching less quickly the weaker $J_2$ is. (c) The Curie law for $J_2 = 0$ slightly differs from that for $J_2 > 0$.   \label{figure15}} 
\end{figure}

The static magnetic susceptibility is one of the first measurements made on new magnetic materials.  It is calculated as,
\begin{equation}
\chi(T) = \frac{{\mathcal{S}}(\vec{q}=\vec{0}, T)}{T} = \frac{1}{N \, T} \displaystyle\sum_{i,j} \! < \! \vec{s_i}(T) \cdot \vec{s_j}(T) \! >\text{.}\label{equation4}
\end{equation}
Plots of the inverse magnetic susceptibility versus temperature are shown in Fig. \ref{figure15}.  At high temperatures (Fig. \ref{figure15} (a)), the inverse susceptibility shows a Curie-Weiss form with a large negative Curie-Weiss constant indicating the dominance of antiferromagnetic couplings between nearest neighboring sites.  A characteristic feature of frustrated magnets is that their magnetic susceptibility shows no signs of magnetic order to well below the Curie-Weiss temperature.  Unlike ordering antiferromagnets which exhibit non-monotonic susceptibilities at the N{\'e}el temperature, for $-J_1\le J_2\le J_1$, the inverse susceptibility monotonically decreases with the temperature.  It is not uncommon for frustrated materials to have defects or uncorrelated ``orphan spins'', such that the susceptibility diverges more quickly at low temperatures, as a Curie law.  In this system we see an intrinsic property of Ising spins on the sorrel lattice is a low temperature Curie tail.

\subsubsection{$J_2 = 0$}

When $J_2 = 0$, the inverse susceptibility crosses over from a high temperature Curie-Weiss law to a low temperature Curie law.  For $L=2$ we find $\chi^{-1} = (3.9134\pm 0.0007) T$.  This likely indicates that a substantial fraction of the spins at the 6-coordinated sites are freely flipping at low temperatures.

\subsubsection{$J_2 > 0$}

For $J_2 > 0$, this is certainly true.  At the lowest temperatures (see Fig. \ref{figure15}(b)) all $J_2 > 0$ curves that we have investigated approach the same Curie law.  For $J_2 = 1.0$, and $L = 4$ we find this to be $\chi^{-1} = (4.000\pm 0.001)T$.  This is consistent with $\frac{1}{4}$ of the spins being completely free Ising spins, while the remaining spins are ordered.  Interestingly, one does see that at temperatures of order $-\theta_{CW} = (5.97\pm0.02)J_1$, the inverse susceptibility begins to deviate upwards from its Curie-Weiss fit, as one might expect from an ordering antiferromagnet.  However, as it begins to order, it frees the 6-coordinated spins, causing $\chi^{-1}$ to eventually downturn at low temperatures once more.  As shown in Fig. \ref{figure15} (a), there is surprisingly little system size-dependence to the magnetic susceptibility, so we have chosen to show $L=2$ plots in Fig. \ref{figure15} (b).  

\subsubsection{$J_2 < 0$}

The low temperature susceptibility for $J_2 < 0$ follows a distinctly different Curie law.  For $L = 2$ it is not hard to write out all possible ground state spin structures consistent with every $J_2$ bond having aligned spins.  So doing, one finds total spins ranging from $S = 0$ to $S=16$ and their degeneracies as shown in Table \ref{tabletwo}.  Writing the magnetization in the limiting case $B\rightarrow 0$ in terms of the magnetic moment per spin state and the partition function allows us to evaluate the magnetic susceptibility and find an exact result, $\chi^{-1}_{L=2} =\frac{32}{14}T\approx 2.2857 T$.  Our Monte Carlo results\cite{notechi} at $L =2 $ agree with this exact calculation for $J_2 = -J_1$ at low temperatures, $\chi^{-1}(T) = (2.28689\pm 0.00059)T - (7\pm 30)\times 10^{-5}$.  The low temperature asymptotes for $J<0$ for higher $L$ appear to agree with this result as well.

\begin{table}
\begin{tabular}{|l|l|}
\hline
Total spin & Degeneracy  \\
\hline
$\pm$16& 1\\
\hline
$\pm$12& 12\\
\hline
$\pm$10&24\\
\hline
$\pm$8&96\\
\hline
$\pm$6&200\\
\hline
$\pm$4&516\\
\hline
$\pm$2&672\\
\hline
0&1054\\
\hline

\end{tabular}
\caption{Degeneracies of the ground state spin configurations satisfying parallel spins on every pair of $J_2$ bonded spins as a function of the total spin of the $L = 2$ spin configuration.  \label{tabletwo} }
\end{table}

\section{Discussion}\label{section5}

\subsection{Non-magnetic depletion}

In this work we have introduced the sorrel net and provided the first study of frustrated magnetism on this depleted triangular lattice.  The regular substitution of non-magnetic atoms into a frustrated magnetic sublattice of a material is a useful way of predicting and perhaps generating new candidate exotic materials.  One such material is the quantum spin liquid candidate Na$_4$Ir$_3$O$_8$\cite{okamoto}, which features $s=\frac{1}{2}$ Ir atoms at three of the four corners of the corner-shared tetrahedral lattice (common to the pyrochlore and B-site spinel geometries).  The Ir atoms in this material are arranged to create a three-dimensional corner-shared triangle hyperkagome lattice.  Non-magnetic Na atoms sit at the fourth corner of each tetrahedron, and whether because of substantial size differences\cite{Shannon} between Na$^+$ and Ir$^{4+}$ ions or electrostatic interactions with surrounding atoms, the Na ions spread throughout the lattice in a regular pattern.  

\subsection{Evidence for depleted triangular lattices}

In this context, it is interesting to note that triangular lattice depletions of the triangular lattice not only yield the ($\frac{1}{9}$th doped) sorrel lattice presented here, but also a number of frustrated lattices for which experimental candidates have already been found. An impressive collection of pictures of possible trigonally symmetric lattices derived by periodic site depletion is presented in the electronic supplementary information to Ref. \onlinecite{keene}. Most notably, the much studied kagome lattice, a $\frac{1}{4}$ doped triangular lattice, has recently been realized experimentally in the materials volborthite\cite{hiroi}, herbertsmithite\cite{shores}, kapellasite\cite{colman}, haydeeite\cite{schluter}, vesignieite\cite{okamotohiroi}, and Cs$_2$Cu$_3$SnF$_{12}$\cite{ono}.  Many of these minerals are Cu based as experimentalists search for low spin quantum spin liquid candidates.  Additionally, artificial spin ice on the kagome lattice was first produced by Qi {\it{et al.}}\cite{Qi} 

 The $\frac{1}{7}$th depleted triangular lattice, the maple leaf lattice, was introduced in 1995 and has candidate materials: M$_x$[Fe(O$_2$CCH$_2$)$_2$NCH$_2$PO$_3$]$_6\cdot n$H$_2$0 where $x =11$ if M is Na or K, and $x =10$ if M is Rb\cite{cave} and spangolite\cite{hawthorne,fennell}.  Even the triangular kagome lattice realized by the Cu atoms\cite{mekata} in Cu$_9$X$_2$(2-carboxypentonic acid)$_6$$\cdot$xH$_2$0, where $X$ is F, Cl or Br, studied by Loh {\it{et al.}}\cite{loh} is a triangular lattice $\frac{7}{16}$th depletion of a triangular lattice, where each point of the triangular lattice removes 7 spins.  

The natural extension of the doping of the triangular lattice from the kagome lattice (which removes every second spin along the reciprocal lattice vectors), to the sorrel lattice (which removes every third spin), would lead to the removal of every fourth spin along the reciprocal lattice vectors for a $\frac{1}{16}$th doping.  Interestingly, for such a lattice it is not possible to remove bonds to create a new corner-sharing triangle lattice.  However, the selective removal of edge-sharing bonds analogous to setting $J_2 = 0$ on the sorrel lattice, would create an interesting potentially frustrated lattice--featuring a star of David decoration inside each empty hexagon of the triangular kagome lattice.

\subsection{A $\frac{1}{9}$th depleted triangular lattice}

Is it possible to create a regular $\frac{1}{9}$th site depleted triangular magnetic lattice in an insulator?  We were pleasantly surprised to learn that this has indeed recently been achieved in a new cobalt hydroxide oxalate Co$_{12}$(OH)$_{18}$(C$_2$O$_4$)$_3$(C$_4$N$_2$H$_{10}$) by Keene {\it{et al.}}\cite{keene}.  In this material the planar Co atoms are octahedrally oxygen coordinated, with $\frac{1}{9}$ of the octahedra empty.  Unfortunately, above and below the empty (in-plane) octahedra are Co atoms.   Three oxygen atoms from the tetrahedra forming the base of a tetrahedral coordination of these Co atoms with nitrogen providing the apical atom of the tetrahedra. It is not immediately clear what effect couplings to such tetrahedral Co atoms might create, although even a weak coupling between the tetrahedral Co atoms might make a strong (ferromagnetic) perturbation on the hexagon of Co atoms nearest to them. Additionally, this layered material has a 2 dimensional honeycomb network of oxalate mediated couplings between Co ions.  Magnetically, this material shows interesting, although perhaps not frustrated behavior, apparently antiferromagnetically ordering at $T_c = 23.5\pm0.5 K$ despite an antiferromagnetic Weiss constant of only about 3.6 K. Given that the authors\cite{keene} expect an antiferromagnetic coupling in the honeycomb layer of 18$\pm 3$K from previous work\cite{andrefwithin}, it seems natural to ask whether or not the bonds of the sorrel layers in this material indeed have antiferromagnetic couplings, or are contributing to lowering the effective Weiss constant. 

One expects that the predominant interaction between Co atoms in the plane is via superexchange through the joining oxygen sites, and that as the Co-O-Co bond varies from 180$^o$ to 90$^o$ at some angle (roughly 120$^o$ in the cuprates) ferromagnetic exchange will be favored over antiferromagnetic exchange.  Of the four in plane bonds presented by the authors, $\{{\mathcal{J}}_1,{\mathcal{J}}_2,{\mathcal{J}}_3,{\mathcal{J}}_4\}$ involving planar Co atoms, our J$_1$ corresponds to the assumption of equal bond strengths on their ${\mathcal{J}}_1$ and ${\mathcal{J}}_2$, our $J_2$ corresponds to their ${\mathcal{J}}_3$ and their ${\mathcal{J}}_4$ is a new bond to the tetragonal Co atoms.  While it would be interesting to carry out a detailed calculation of these exchange integrals in future work, na{\"i}vely from bond angle considerations\cite{bondangles} one would say that ${\mathcal{J}}_4$  is likely to be weakly antiferromagnetic,  ${\mathcal{J}}_1$ and ${\mathcal{J}}_2$  may well be weakly ferromagnetic, and ${\mathcal{J}}_3$  is likely to be weakly ferromagnetic. As such, one might expect that this system falls in a regime not covered in this work, of ferromagnetic $J_1$ and $J_2$.   It would certainly also be interesting if a local probe such as oxygen NMR could be used to determine the signs of the magnetic couplings within the plane.  If, on the other hand, it were shown that this material possesses an antiferromagnetic $J_1$, one might hope to exploit the differences between the atomic sizes of octahedrally and tetrahedrally coordinated Co$^{2+}$ ions to design a non-magnetic substitution.  For example, according to Shannon\cite{shannon1}, tetrahedrally coordinated Mg$^{2+}$ is very close in size to tetrahedrally coordinated Co$^{2+}$, while the octahedral coordinated ions of both species are of quite different sizes, suggesting that if it were possible to substitute Mg atoms for Co atoms, the Mg atoms might preferentially enter the tetrahedral site, removing some complicating non-frustrated magnetic physics.

\subsection{Designing for antiferromagnetic superexchange}

While we view the solution of the Ising model on the sorrel lattice as a simple test whether or not antiferromagnetism would be frustrated on this net, it is interesting to speculate on how this, or related models might be realized in an experimental system.  In particular, one might wonder how magnetic exchange processes could lead to an antiferromagnetic $J_1$ and a weak, non-existent, or ferromagnetic $J_2$, even if the magnetic atoms of a crystal could be arranged into the $\frac{1}{9}$th doped triangular net.  One could imagine\cite{possible} superexchange pathways via say oxygen atoms living at the midpoints of the $J_1$ bonds leading to a strong antiferromagnetic coupling on these bonds alone.  Indeed even oxygen atoms lying at the center of the $J_1$ triangles would likely lead to weakly antiferromagnetic $J_1$ bonds.  Such interactions might be expected to produce at least roughly equal strength bonds around both the 6-coordinated and 4-coordinated spin sites, as we have assumed in this work.  To additionally realize Ising spins would likely require strong crystal field anisotropies in the out of plane direction, leading to a magnetically easy axis.

\subsection{A two-dimensional structure}

That the sorrel net is two-dimensional opens a host of possible realizations for further study.  For example, recent advances in the study of spinless bosons trapped by optical lattices have allowed the simulation of models of frustrated magnetism on the triangular lattice with couplings of various magnitudes.\cite{struck}  Could such studies be extended to the doped triangular lattices?  From soft condensed matter, the study of magnetic colloids interacting via the dipolar interaction offers the possibility of real time manipulation of a lattice using optical tweezers to perhaps create a mesoscopic frustrated lattice of choice.{\cite{steinbach}}

While it may be currently unrealistic\cite{nogo} to expect the creation of a direct artificial realization of a dipolar Ising model on the sorrel lattice with moments pointing out of the plane,  the construction of a coplanar artificial spin ice structure should be feasible.  As for (lithographically etched) realizations of artificial spin ice on the kagome lattice\cite{}, a permalloy island could be to chosen to point locally along the axis of symmetry of the triangle.  While such islands could join the centers of the triangles about the 4-coordinated spins sites, a new and interesting feature for study would arise at the 6-coordinated spin sites.  Three overlapping ferromagnetic islands (with a $C_6$ symmetry) would mimic a multiorbital local  Ising spin at these vertices, producing a novel interplay between short range contact terms (Hund's coupling and ferromagnetic interactions) with the long range dipolar interaction.  We have begun \cite{ianandme} the study of ferromagnetic interactions between such local Ising spins in the presence of Hund's coupling terms and find, at least to this approximation, that degeneracy remains in the ground state of the sorrel lattice.  Unfortunately, this multiorbital nature appears to prevent the creation of a macroscopic spin ice along the lines of that created by Mellado {\it{et al.}}\cite{melladomacro}

The question of what quantum magnetism will do on the sorrel lattice is left for future work.   When considering frustrated quantum spin systems, perhaps the simplest model is that of the quantum transverse Ising model\cite{moessner}, the study of which is considerably simplified for low dimensional lattices.
  We\cite{jarrettandme} have begun to investigate classical dimer models on the sorrel lattice using the Pfaffian approach.  The hope is that as on the triangular lattice, the classical and quantum dimer models may be related, and there may be a mapping\cite{moessner} to the quantum transverse Ising model.

%

\section{Conclusions}

We have studied the antiferromagnetic Ising model on a new corner-shared equilateral triangle net, finding a large finite residual entropy $\frac{S}{N}$ =0.48185$\pm$0.00008, slightly larger than the corresponding Pauling entropy $\frac{S}{N}=\frac{1}{4}\ln(\frac{27}{4})\approx 0.477386$.  We have shown that the spin-spin correlations  remain disordered to the lowest temperature, with the structure factor showing broad disperse peaks about $\vec{q} = (\frac{4\pi}{a},0)$ and $C_6$ rotations.  The magnetic susceptibility indicates that strong antiferromagnetic correlations at high temperatures give way to short range ordering at low temperatures, with a Curie tail with slope close to that one would expect were $\frac{1}{4}$ of the Ising spins completely free.  While we have not proposed an exact solution, our results are remarkably independent of system size, indicating that the thermodynamic limit is quickly reached.  

To this corner-shared triangle sorrel net we have added edge-sharing bonds ($J_2$) which are likely to remain present in physical realizations of the doped triangular lattice, and investigated the phase diagram produced by varying the magnitude and sign of this coupling constant.  

For antiferromagnetic edge-sharing couplings ($J_2 > 0$), we have shown that at low temperatures our spin system adopts a partially ordered state with Ising spins at the 6-coordinated sites completely free, and all other neighboring spins antiferromagnetically correlated, resulting in sharp magnetic Bragg peaks at $\vec{q} = (\frac{4\pi}{a}, 0)$ and $C_6$ rotations.  This allows us to write exact solutions for the ground state entropy $\frac{S}{N} = (\frac{1}{4}+\frac{1}{8L^2})\ln(2)$ and the low temperature Curie tail of the susceptibility, results which are supported numerically.

For ferromagnetic edge-sharing couplings ($J_2<0$), we have shown that at sufficiently low temperatures all combinations of ferromagnetically aligned spins across edge-sharing ($J_2$) bonds are equally realized, resulting in diffuse magnetic scattering about $\vec{q} = (\frac{4\pi}{a},0)$ and $C_6$ rotations, similar to the pure sorrel net case.  In contrast to this pure case, we have shown that the low temperature susceptibility shows a considerably stronger Curie tail, which we have found to be consistent with exact $L=2$ results.  Knowing the ground state, we have found an exact expression for the residual entropy  $\frac{S}{N} = \frac{3\ln(2)}{8}$ of the spin system which is supported numerically. 

\section{Acknowledgments}

This work was supported by NSERC (JMH), NSERC USRA (JJB), the Brandon University Research Committee (JJB) and Manitoba Career Focus (JJB).  JMH would like to thank Travis Redpath, John Chalker, Roderich Moessner, Chris Henley and Onofre Rojas for useful comments during the completion of this manuscript.


\begin{thebibliography}{00}
\bibitem{henley} We would like to thank Chris Henley for pointing out that we should make this point.
\bibitem{note3} For calculations of the entropy, Monte Carlo simulations were run from a starting temperature of $6000|J|$ down to $9.99\times10^{-4}|J|$.
\bibitem{notechi} This result for the slope is within 2 standard deviations of the exact result and corresponds to $10^6$ MCS, and the heat capacity as presented in this work.  With only $2\times 10^5$ MCS and the same seeds, the same low temperature fit of the susceptibility (from $T = 0.1583 J_1$ to the lowest simulated temperature) yielded $\chi^{-1} = (2.2857\pm 0.0006)T + (7 \times 10^{-6}\pm3\times 10^{-5})$.  In both cases, the intercept is zero.
\bibitem{Shannon} According to Ref \onlinecite{shannon1}, the crystal radii (in $\AA$) of Ir$^{4+}$ and Na$^+$ in an octahedral environment are respectively 0.765 and 1.16.
\bibitem{possible} If only magnetic atoms and oxygen atoms were involved, it is not immediately clear to us what oxygen coordination the magnetic atoms would end up with in such a scenario.
\end{thebibliography}
\end{document}